\documentclass[intlimits,sumlimits,12pt]{iopart}

\newcommand{\diag}{\textrm{diag}}
\newcommand{\dd}{\text{d}}
\newcommand{\trans}{\text{T}}
\newcommand{\ytens}{\left(Y\otimes \mathds{1}_N\right)}
\newcommand{\ytenstrans}{\left(Y^\trans \otimes \mathds{1}_N\right)}

\newcommand{\bmk}{\bm{k}}
\newcommand{\bmkstar}{\bm{k}^\star}
\newcommand{\absk}{\vert\bm{k}\vert}

\usepackage{iopams}
\expandafter\let\csname equation*\endcsname\relax

\expandafter\let\csname endequation*\endcsname\relax

\usepackage{amsmath}
\usepackage{hyperref}
\usepackage{dsfont}
\usepackage{cite}
\usepackage{graphicx}
\usepackage{amssymb}
\usepackage{mathtools}
\usepackage[capitalise]{cleveref}
\crefname{section}{Sec.}{Secs.}
\usepackage{braket}
\usepackage[disable]{todonotes}
\usepackage{float}
\usepackage{subfig}
\usepackage{nicematrix}
\usepackage{bm}
\NiceMatrixOptions{cell-space-limits = 2pt}

\DeclareMathOperator{\str}{str}
\DeclareMathOperator{\sdet}{sdet}

\newcommand{\tauthree}{\tau^{(3)}}
\newcommand{\fourotau}{\left(\mathds{1}_4 \otimes \tauthree\right)}
\newcommand{\twootau}{\left(\mathds{1}_2 \otimes \tauthree\right)}

\newcommand*{\myprime}{^{\prime}\mkern-1.2mu}

\newcommand{\bralign}[1]{\phantom{#1} \quad}

\allowdisplaybreaks
\begin{document}

\title[]{Influence of Time Reversal Invariance Breaking on the Distribution of Off-Diagonal Scattering Matrix Elements and Cross Sections}

\author{Nils Gluth, Ahmed Aldabag and Thomas Guhr}

\address{Fakult\"at f\"ur Physik, Universit\"at Duisburg--Essen, Duisburg, Germany}
\ead{nils.gluth@uni-due.de, ahmed.aldabag@stud.uni-due.de, thomas.guhr@uni-due.de}
\vspace{10pt}

\begin{abstract}
  Scattering theory is a key tool for the investigation of quantum systems. Universal approaches are of interest, particularly in the framework of Random Matrix Theory. Dyson's Threefold way classifies systems according to their behavior under time reversal. In these three cases the distribution of off-diagonal scattering matrix elements and cross sections were recently calculated. However, in many applications systems neither display full nor fully broken time reversal invariance, such as systems in tuneable external magnetic fields. The study of these systems has a long history, also in Random Matrix Theory. We investigate the influence of time reversal invariance breaking on the universal behavior of the mentioned distributions. 
Using supersymmetry we succeed in deriving exact expressions for the distribution of off-diagonal scattering matrix elements and cross sections in systems with half-integer or integer/no spin in the absence of further symmetries. We find that the distributions are symmetric while the scattering matrix elements are not. Furthermore, our results might considerably facilitate the data analysis of time reversal invariance breaking.
\end{abstract}

\vspace{2pc}

\submitto{}

\section{Introduction}
\label{sec0}
This paper is the last in a series of three papers \cite{GG2025,GG2025_2}. We refer the reader to \cite{GG2025} for an introduction to the topic as well as to some of the salient features and to the literature. As it is particularly relevant here we give the definition of the three Dyson classes \cite{Dyson1962a} labelled by the Dyson index $\beta$
\begin{enumerate}
	\item Gaussian Orthogonal Ensemble (GOE, $\beta=1$) \\
	Time reversal invariant systems with rotational symmetry and systems with integer spin without further symmetries. These systems are governed by a real symmetric Hamilton operator
	\begin{equation}
		H = H^\star = H^\trans
	\end{equation}
	\item Gaussian Unitary Ensemble (GUE, $\beta=2$) \\
	Systems without time reversal invariance. The Hamilton operator is complex Hermitian
	\begin{equation}
		H = H^\dagger
	\end{equation}
	\item Gaussian Symplectic Ensemble (GSE, $\beta=4$) \\
	Time reversal invariant systems with half-integer spin without further symmetries. The Hamilton operator is real quaternion self-dual
	\begin{equation}
		H = H^\dagger, \quad H = \begin{bNiceMatrix}
			H_0 & H_1 \\
			-H_1^\star & H_0^\star
		\end{bNiceMatrix} .
	\end{equation}
\end{enumerate}
The paper is structured as follows. We introduce time reversal invariance breaking to our random matrix model in \cref{sec:timeRevInvBreaking}. Furthermore, we discuss the two distinct cases of broken and unbroken Kramers' degeneracy for GSE systems. We also recapitulate the historic relevance of time reversal invariance testing and argue that our results allow a more detailed analysis of experiments than what has been possible before. Then we derive an exact result for the distribution of scattering matrix elements using supersymmetry for the GSE in \cref{sec:gsegueScatterMatElem} and the GOE in \cref{sec:goegueScatterMat}. Finally, we discuss our result in \cref{sec:conclusion}. 

\section{Breaking Time Reversal Invariance}
\label{sec:timeRevInvBreaking}
In \cref{subsec:ChoiceHamilton}, we introduce our choice of transitional ensemble by which we model time reversal invariance breaking. We formulate this for both the GOE and GSE. In the latter case we also discuss the role of Kramers' degeneracy. Following this, we detail the historic importance of detailed balance experiments and the limits of past data analysis in \cref{subsec:DetailedBalance}.
\subsection{Choice of Hamiltonian and Parametrization}
\label{subsec:ChoiceHamilton}
In GOE systems time reversal invariance breaking is usually modelled by \cite{MP1983}
\begin{equation}\label{eqn:goeIB}
	H = H_0 + i \alpha H_1
\end{equation}
with a real symmetric $H_0$, a real skew-symmetric $H_1$ and a transition parameter $\alpha$. Clearly, for $\alpha=0$ we recover the real symmetric case and for $\alpha=1$ we obtain the Hermitian case. This parametrization was used in Ref. \cite{PWZLW1995} to calculate correlations of scattering matrix elements. We use the same model in our subsequent calculations.

In contrast to system with orthogonal symmetry, the structure of the Hamiltonian due to the half-integer spin in GSE systems allows for two distinct types of transitions. The first possibility is to preserve the Kramers' degeneracy \cite{K1930} such that
\begin{equation}
	H = \begin{bNiceMatrix}
		H_0 & 0 \\
		0 & H_0^\star
	\end{bNiceMatrix} + \alpha \begin{bNiceMatrix}
	0 & H_1 \\
	- H_1^\star & 0
	\end{bNiceMatrix}
\end{equation}
with a Hermitian matrix $H_0$, a complex skew-symmetric $H_1$ and a transition parameter $\alpha$. In this case we find a Hermitian matrix with Kramers' degeneracy for $\alpha=0$ and a self-dual matrix for $\alpha=0$. Alternatively, the second possibility is to break Kramers' degeneracy as well as the time reversal invariance by
\begin{equation}
	H = H_{\text{GSE}} + i \alpha H_{\text{skewGSE}}
\end{equation}
with a self-dual matrix $H_{\text{GSE}}$ and a skew-Hermitian $H_{\text{skewGSE}}$ with symplectic symmetry, \textit{i.e.}
\begin{equation}
	H_{\text{skewGSE}} = \begin{bNiceMatrix}
		V_0 & V_1 \\
		-V_1^\star & V_0^\star
	\end{bNiceMatrix} .
\end{equation}
Here we find the time reversal invariant case for $\alpha=0$ and fully break the invariance for $\alpha=1$ yielding a Hermitian matrix without any further structure, but with twice the usual dimension.

In the present work we will only focus on the case disregarding Kramers' degeneracy as it is closely related to a forthcoming study \cite{GG2025}. Anticipating a later saddle point approximation we set 
\begin{equation}
	\alpha =\pi\xi/\sqrt{N} .
\end{equation}
Preserving Kramers degeneracy while breaking time reversal invariance is still interesting regardless as it is closely related to some experimental setups of symplectic quantum graphs \cite{CGKGD2025}. However, it requires different supersymmetric treatment and is part of ongoing research that goes beyond the scope of this work.

\subsection{Detailed Balance Experiments}
\label{subsec:DetailedBalance}
The experimental study of time reversal invariance breaking has a rich history in nuclear scattering dating back to the 70s. As a consequence of time reversal invariance in systems with integer spin or rotational symmetry, the scattering amplitudes $S_{ab}$ and $S_{ab}$ must equal. This directly implies a detailed balance $\vert S_{ab}\vert^2 = \vert S_{ba}\vert^2$ of the reactions $a\to b$ and $b\to a$. Testing the existence of detailed balance has been the main tool to investigate time reversal invariance. For further information we refer the reader to Ref. \cite{MRW2010} and only list some of the important points here. Some of the early tests of detailed balance like Ref. \cite{DBGRSP1979,BDGGRS1983} were carried out by comparing the ratio of experimentally obtained cross sections and reaction rates for the two possible reactions to one. The result is an asymmetry parameter $\delta$ where $\delta\sim0$ indicates detailed balance. These experiments found that time reversal invariance is not broken by the strong interaction. Furthermore, they improved already existent upper limits for the possible strength of time reversal invariance breaking. Interestingly, dating back to Ref. \cite{HJ1959}, detailed balance does not imply time reversal invariance and there exist systems in which $S_{ab} = S_{ba}$ but time reversal invariance is broken. Importantly, this is actually the case for GUE systems, see the results of \cite{KNSGDMRS2013}. Thus, instead of testing for detailed balance a statistical analysis of the scattering matrix elements and cross sections gives better tools to test for time reversal invariance breaking. Such an analysis was for example carried out in Ref. \cite{DFH2009} with the help of microwave networks in which the authors used analytically derived cross-correlations to test for time reversal invariance violations. In our recent works we managed to exactly derive the distributions of real and imaginary parts and cross sections of scattering matrix elements for all three Dyson classes \cite{KNSGDMRS2013,NKSG2014,GDG2025,GG2025} which provide a new and more sophisticated avenue for analysis of the experimental data. Hence, our goal is to also derive such expressions for the ensembles which allow time reversal invariance breaking.

\section{Time Reversal Invariance Breaking for Symplectic Symmetry}
\label{sec:gsegueScatterMatElem}
In \cref{subsec:DistributionGSE}, we introduce the distribution of the off-diagonal scattering matrix elements. Then we perform the ensemble average in \cref{subsec:EnsembleAverage}. After identifying an appropiate choice of supermatrix, we carry out the integration over a real supervector in \cref{subsec:RealSupervector} and subsequently perform a saddle point approximation in \cref{subsec:SaddlePoint}. We integrate over the saddle point manifold and discuss the resulting expression in \cref{subsec:IntegrationSaddlePoint}. Concluding this sections, we present results for the distribution of the scattering cross sections and briefly discuss the calculation of correlators in \cref{subsec:CrossSectionsGSE}.
\subsection{Distribution of Off-Diagonal Scattering Matrix Elements}
\label{subsec:DistributionGSE}
Many of the steps necessary to calculate the characteristic function are the same as in Ref. \cite{GG2025}. Hence, we focus on the necessary modifications of the calculations, without recapitulating the details.

We want to calculate the distribution of real $(s=1)$ and imaginary $(s=2)$ part of the scattering matrix elements
\begin{equation}
	\wp_s\left(S_{am bm\myprime}(E)\right) = \frac{1}{2i^{s-1}} \left(S_{am bm\myprime}(E) + (-1)^{s-1} S_{am bm\myprime}^\star(E)\right)
\end{equation}
via an ensemble average of a filter function
\begin{equation}
	P_{s,m m\myprime}(x_s\vert\xi) = \int\dd[H] \mathcal{P}(H) \delta\left(x_s - \wp_s\left(S_{am bm\myprime}(E)\right)\right)
\end{equation}
where the distribution in this case is 
\begin{align}\label{eqn:defPH}
	\mathcal{P}(H) =& \int\dd[H_{\text{GSE}}] \int\dd[H_{\text{skewGSE}}] \delta\left(H - \left(H_{\text{GSE}} + \frac{i \pi\xi}{\sqrt{N}} H_{\text{skewGSE}}\right)\right)\notag\\
	&\times \mathcal{P}^{(4)}(H_{\text{GSE}}) \mathcal{P}^{(4)}_{\text{skew}}\left(H_{\text{skewGSE}}\right) , \notag\\
	\mathcal{P}^{(4)}(H_{\text{GSE}}) \sim& \exp\left(-\frac{N}{2v^2} \tr H_{\text{GSE}}^2\right), \notag\\
	\mathcal{P}^{(4)}_{\text{skew}}\left(H_{\text{skewGSE}}\right) \sim& \exp\left(+\frac{N}{2v^2} \tr H_{\text{skewGSE}}^2\right)
\end{align}
where the plus sign in the exponent appears for skew-Hermitian matrix as their squares are negative semi-definite. Equivalently, starting from the characteristic function,
\begin{equation}
	R_{s,m m\myprime}(k\vert \xi) = \int\dd[H] \mathcal{P}(H) \exp\left(- i k \wp_s\left(S_{am bm\myprime}(E)\right)\right),
\end{equation}
replacing the phase by a Gaussian integration over commuting and anticommuting variables, and applying a transformation we find
\begin{align}
	R_{s,m m\myprime}(k\vert \xi) =& \int\dd[\psi] \exp\left(\frac{i}{2} \left(\psi^\dagger \widetilde{\mathbf{W}} + \widetilde{\mathbf{U}}_s^\dagger \psi\right)\right) \int\dd[H] \mathcal{P}(H) \exp\left(\frac{i}{4\pi k} \psi^\dagger \mathbf{\Omega} \psi\right) , \notag\\
	\psi =& \left(z_a, z_b, \zeta_a, \zeta_b\right), \quad \mathbf{\Omega} = \diag\left(- \left(G^{-1}\right)^\dagger, G^{-1}, - \left(G^{-1}\right)^\dagger, - G^{-1}\right), \notag\\
	\widetilde{\mathbf{W}} =& \left(\widetilde{W}, 0\right), \quad \widetilde{\mathbf{U}}_s^\dagger = \widetilde{\mathbf{W}}^\dagger \left(\Xi^+ \oplus \Xi^-\right)
\end{align}
with
\begin{align}
	\widetilde{W} =& \begin{cases}
		\left(\widetilde{W}_a, \widetilde{W}_b\right) &, m = \uparrow, m\myprime = \uparrow \vspace{0.1cm}\\
		\left(\widetilde{W}_a, \ytens\widetilde{W}_b^\star\right) &, m = \uparrow, m\myprime = \downarrow \vspace{0.1cm}\\
		\left(\ytens \widetilde{W}_a^\star, \widetilde{W}_b\right) &, m = \downarrow, m\myprime = \uparrow \vspace{0.1cm}\\
		\left(\ytens \widetilde{W}_a^\star, \ytens \widetilde{W}_b^\star\right) &, m = \downarrow, m\myprime = \downarrow
	\end{cases}, \notag\\
	\Xi^\pm =& \begin{bNiceMatrix}
	0 & \pm\left(-i\right)^s \mathds{1}_{2N} \\
	-i^s \mathds{1}_{2N} & 0
	\end{bNiceMatrix} .
\end{align}
We separate the matrix $\mathbf{\Omega}$ into parts dependent and not depended on $H$
\begin{align}
	\psi^\dagger \mathbf{\Omega} \psi =& \tr H K + \psi^\dagger \mathbf{\Omega}_0 \psi, \notag\\
	 K =& z_a z_a^\dagger - z_b z_b^\dagger - \zeta_a \zeta_a^\dagger - \zeta_b \zeta_b^\dagger , \notag\\
	 \mathbf{\Omega}_0 =& E \diag\left(-1,1,-1,-1\right) \otimes \mathds{1}_{2N} + i \pi \diag\left(1,1,1,-1\right) \otimes \sum_{c=1}^{M} W_c W_c^\dagger
\end{align}
such that
\begin{align}
	R_{s,m m\myprime}(k\vert \xi) =& \int\dd[\psi] \exp\left(\frac{i}{2} \left(\psi^\dagger \widetilde{\mathbf{W}} + \widetilde{\mathbf{U}}_s^\dagger \psi\right)\right) \exp\left(\frac{i}{4\pi k} \psi^\dagger \mathbf{\Omega}_0 \psi\right) \notag\\
	&\times \int\dd[H] \mathcal{P}(H) \exp\left(\frac{i}{4\pi k} \tr H K\right) .
\end{align}
Inserting the definition of the distribution $\mathcal{P}(H)$, \textit{cf.} \cref{eqn:defPH}, the ensemble average factorizes into an integration over $H_{\text{GSE}}$ and $H_{\text{skewGSE}}$
\begin{align}
	&\int\dd[H] \mathcal{P}(H) \exp\left(\frac{i}{4\pi k} \tr H K\right) \notag\\
	=& \int\dd[H_{\text{GSE}}] \mathcal{P}^{(4)}(H_{\text{GSE}}) \exp\left(\frac{i}{4\pi k} \tr H_{\text{GSE}} K\right) \notag\\
	&\times \int\dd[H_{\text{skewGSE}}] \mathcal{P}_{\text{skew}}^{(4)}(H_{\text{skewGSE}}) \exp\left(\frac{i}{4\pi k} \tr \frac{i \pi \xi}{\sqrt{N}} H_{\text{skewGSE}} K\right) .
\end{align}

\subsection{Performing the Ensemble Average}
\label{subsec:EnsembleAverage}
The first integration was already performed in Ref. \cite{GG2025} and yields
\begin{align}\label{eqn:ensembleAverageGSE}
	\int\dd[H_{\text{GSE}}] \mathcal{P}^{(4)}(H_{\text{GSE}}) &\exp\left(\frac{i}{4\pi k} \tr H_{\text{GSE}} K\right) = \exp\left(-\frac{v^2}{2N (8\pi k)^2} \tr\widehat{K}_+^2\right), \notag\\
	\widehat{K}_+ =& K + \ytenstrans K^\trans \ytens, 
\end{align}
where $Y=-i\tau^{(2)}$ is up to a prefactor the second Pauli matrix $\tau^{(2)}$.
We perform the second integration in the same fashion and arrive at a very similar result
\begin{align}\label{eqn:ensembleAverageskewGSE}
	\int\dd[H_{\text{skewGSE}}] \mathcal{P}_{\text{skew}}^{(4)}(H_{\text{skewGSE}})& \exp\left(\frac{i}{4\pi k} \tr \frac{i \pi \xi}{\sqrt{N}} H_{\text{skewGSE}} K\right) = \exp\left(- \frac{\pi^2\xi^2 v^2}{2N^2(8\pi k)^2} \tr\widehat{K}_-^2\right), \notag\\
	\widehat{K}_- =& K - \ytenstrans K^\trans \ytens .
\end{align}
Similar to the symplectic case, \textit{c.f.} Ref. \cite{GG2025}, we express the matrices $\widehat{K}_\pm$ as a product of the supermatrix
\begin{equation}
	A = \begin{bNiceMatrix}
		z_a^\dagger \\
		z_a^\trans \ytenstrans \\
		z_b^\dagger \\
		z_b^\trans \ytenstrans \\
		-\zeta_a^\dagger \\
		\zeta_a^\trans \ytenstrans \\
		-\zeta_b^\dagger \\
		\zeta_b^\trans \ytenstrans
	\end{bNiceMatrix}
\end{equation}
and its conjugate together with a metric
\begin{align}
	\widehat{K}_+ =& A^\dagger \widetilde{L} A, \quad \widetilde{L} = \diag\left(1,-1,1,1\right) \otimes \mathds{1}_2, \notag\\
	\widehat{K}_- =& A^\dagger \left(\mathds{1}_4 \otimes \tauthree\right) \widetilde{L} A .
\end{align} 
Additionally, we write the traces in \cref{eqn:ensembleAverageGSE,eqn:ensembleAverageskewGSE} in terms of supertraces \cite{Guhr2006}
\begin{align}
	\tr\widehat{K}_+^2 =& \str B^2, \quad B = \widehat{L}^{1/2} A A^\dagger \widehat{L}^{1/2}, \notag\\
	\tr\widehat{K}_-^2 =& \str \left(\left(\mathds{1}_4 \otimes \tauthree\right) B\right)^2.
\end{align}
Hence, the full ensemble average is 
\begin{equation}
	\int\dd[H] \mathcal{P}(H) \exp\left(\frac{i}{4\pi k}\tr H K\right) = \exp\left(-\frac{v^2}{2N (8\pi k)^2} \str \left(B^2 + \frac{\pi^2 \xi^2}{N} \left(\left(\mathds{1}_4 \otimes \tauthree\right) B\right)^2\right)\right) .
\end{equation}
Employing another Fourier transform, now in superspace, we have
\begin{align}\label{eqn:ftsuperspace}
	&\exp\left(-\frac{v^2}{2N (8\pi k)^2} \str \left(B^2 + \frac{\pi^2 \xi^2}{N} \left(\left(\mathds{1}_4 \otimes \tauthree\right) B\right)^2\right)\right) \notag\\
	=& \mathcal{N}\int\dd[\sigma] \exp\left(-N r \str \left\{\sigma^2 + \frac{\pi^2 \xi^2}{N} \left( \fourotau \sigma\right)^2\right\}\right) \notag\\
	&\times \exp\left(i \str \left(\sigma + \frac{\pi^2 \xi^2}{N} \fourotau \sigma \fourotau\right) B\right),
\end{align}
where $r=(8\pi k)^2/(2v^2)$. The integration as such is not convergent, but introducing the same convergence generating terms as in Ref. \cite{GG2025} solves this problem. As these terms do not influence the final result we choose to not explicitly write them out and refer the reader to Ref. \cite{GG2025}. Furthermore, the supermatrix $\sigma$ has to inherit the symmetries of $B$ which is guaranteed by 
\begin{align}\label{eqn:gseSigmaDiag}
	\sigma =& T^{-1} \mathfrak{V} \diag\left(\sigma_{\text{B},1}\mathds{1}_2, \sigma_{\text{B},2}\mathds{1}_2,i\sigma_{\text{F},1},i\sigma_{\text{F},2},i\sigma_{\text{F},3},i\sigma_{\text{F},4}\right) \mathfrak{V}^\dagger T, \notag\\
	T^\dagger \widetilde{L} T =& \widetilde{L}, \quad C T^\star C^\trans = T, \quad C = \diag\left(Y, Y^\trans, X, X\right), \notag\\
	X=&\tau^{(1)}, \quad \mathfrak{V}= \diag\left(\mathds{1}_4,\mathds{1}_2\otimes\mathfrak{v}\right), \quad \mathfrak{v}=\frac{1}{\sqrt{2}} \begin{bNiceMatrix}
		1 & i \\
		1 & -i
	\end{bNiceMatrix} .
\end{align}
The transformations $T$ form the non-compact unitary orthosymplectic supergroup $\text{UOSp}(2,2\vert 4)$. For future convenience we introduce
\begin{equation}
	\sigma(\xi) = \sigma + \frac{\pi^2 \xi^2}{N} \fourotau \sigma \fourotau .
\end{equation}
As a consequence of the Fourier transformation in \cref{eqn:ftsuperspace} we are able to carry out the integration over the elements of $\psi$. 
\subsection{Integration over Real Supervectors}
\label{subsec:RealSupervector}
We rewrite the supertrace involving $\sigma(\xi)$ and $B$ in terms of a quadratic form
\begin{equation}
	\str \sigma(\xi) B = \Psi^\dagger \left(\widetilde{L}^{1/2} \sigma(\xi) \widetilde{L}^{1/2}\right)\otimes \mathds{1}_{2N} \Psi
\end{equation}
where the $16N$-dimensional supervector contains the columns of $A^\dagger$
\begin{equation}
	\Psi = \left(z_a, \ytens z_a^\star, z_b, \ytens z_b^\star, \zeta_a, \ytens \zeta_a^\star, \zeta_b, \ytens z_b^\star\right). 
\end{equation}
Expressing the rest of the integral in terms of $\Psi$ gives
\begin{equation}
	\widetilde{\bm{U}}_s^\trans \psi + \psi^\dagger \widetilde{\bm{W}} = \widetilde{\bm{V}}_s^\trans \left(\mathds{1}_4 \otimes \diag\left(\mathds{1}_{2N}, \ytenstrans\right)\right) \Psi
\end{equation}
with the supervector $\widetilde{\bm{V}}_s^\trans = \begin{bNiceMatrix}
	\widetilde{V}_s^\trans & 0
\end{bNiceMatrix}$,
\begin{align}
	\widetilde{V}_{s,\uparrow\uparrow}^\trans =& \begin{bNiceMatrix}
		-i^s \widetilde{W}_b^\dagger & \widetilde{W}_a^\trans & (-i^s)\widetilde{W}_a^\dagger & \widetilde{W}_b^\trans 
	\end{bNiceMatrix} \notag\\
	\widetilde{V}_{s,\uparrow\downarrow}^\trans =& \begin{bNiceMatrix}
		-i^s \widetilde{W}_b^\trans \ytenstrans & \widetilde{W}_a^\trans & (-i^s)\widetilde{W}_a^\dagger & \widetilde{W}_b^\dagger \ytenstrans 
	\end{bNiceMatrix} \notag\\
	\widetilde{V}_{s,\downarrow\uparrow}^\trans =& \begin{bNiceMatrix}
		-i^s \widetilde{W}_b^\dagger & \widetilde{W}_a^\dagger \ytenstrans & (-i^s)\widetilde{W}_a^\trans \ytenstrans & \widetilde{W}_b^\trans 
	\end{bNiceMatrix} \notag\\
	\widetilde{V}_{s,\downarrow\downarrow}^\trans =& \begin{bNiceMatrix}
		-i^s \widetilde{W}_b^\trans \ytenstrans & \widetilde{W}_a^\dagger \ytenstrans & (-i^s)\widetilde{W}_a^\trans \ytenstrans & \widetilde{W}_b^\dagger \ytenstrans 
	\end{bNiceMatrix} .
\end{align}
Furthermore, the symplectic properties of $W_c W_c\dagger$ yield
\begin{align}
	\frac{1}{2} \Psi^\dagger \bm{\mathcal{A}}_0^{-1} \Psi =& \psi^\dagger \bm{\Omega}_0 \psi, \notag\\
	\bm{\mathcal{A}}_0^{-1} = - E \widetilde{L} \otimes \mathds{1}_{2N} + i \pi \widetilde{L} L \otimes \sum_{c=1}^{M} W_c W_c^\dagger&, \quad L = \diag\left(1,-1,1,-1\right) \otimes \mathds{1}_2 .
\end{align}
We return to the characteristic function and with all of the above steps have
\begin{align}\label{eqn:cfbeforesupervectorintegration}
	R_{s,m m\myprime}(k\vert \xi) =& \mathcal{N}\int\dd[\sigma] \exp\left(- N r \str \left\{\sigma^2 + \frac{\pi^2 \xi^2}{N} \left(\fourotau \sigma\right)^2\right\}\right) \notag\\
	&\times \int\dd[\psi] \exp\left(\frac{i}{2} \widetilde{\bm{V}}_s^\trans \left(\mathds{1}_4 \otimes \diag\left(\mathds{1}_{2N}, \ytenstrans\right)\right) \Psi\right) \notag\\
	&\times \exp\left(i \Psi^\dagger \left(\widetilde{L}^{1/2} \sigma(\xi) \widetilde{L}^{1/2} \otimes \mathds{1}_{2N} + \frac{1}{8\pi k}\bm{\mathcal{A}}_0^{-1}\right) \Psi\right) .
\end{align}
Similar to Ref. \cite{GG2025} the integration over the supervector $\psi$ is carried out via a mapping onto a real supervector and decoupling the integration over commuting anticommuting variables. The procedure requires that the Gaussian like term in \cref{eqn:cfbeforesupervectorintegration} fulfills certain symmetries. It is known that the time reversal invariant terms fulfill this symmetry and straightforward calculations show that this is also true for the invariance breaking term. Hence, the integral over the supervector amounts to
\begin{align}\label{eqn:intRealSupervector}
	&\int\dd[\psi] \exp\left(\frac{i}{2} \widetilde{\bm{V}}_s^\trans \left(\mathds{1}_4 \otimes \diag\left(\mathds{1}_{2N}, \ytenstrans\right)\right) \Psi\right) \notag\\
	&\times \exp\left(i \Psi^\dagger \left(\widetilde{L}^{1/2} \sigma(\xi) \widetilde{L}^{1/2} \otimes \mathds{1}_{2N} + \frac{1}{8\pi k}\bm{\mathcal{A}}_0^{-1}\right) \Psi\right) \notag\\
	=& \sdet^{-1/2} \Sigma \exp\left(-\frac{i}{16} \widehat{\bm{V}}_s^\trans \left(\widetilde{L}^{1/2} \otimes\mathds{1}_{2N}\right) \Sigma^{-1} \left(\widetilde{L}^{1/2} \otimes\mathds{1}_{2N}\right) \overline{\bm{V}}_s\right), \notag\\
	\Sigma =& \sigma_E(\xi) \otimes\mathds{1}_{2N} + \frac{i}{8k} L \otimes \sum_{c=1}^{M} W_c W_c^\dagger, \quad \sigma_E(\xi) = \sigma(\xi) - \frac{E}{8\pi k} \mathds{1}_8, \notag\\
	\widehat{\bm{V}}_s^\trans =& \widetilde{\bm{V}}_s^\trans \left(\mathds{1}_4 \otimes \diag\left(\mathds{1}_{2N}, \ytenstrans\right)\right) \notag\\
	\overline{\bm{V}}_s =& \left(\mathds{1}_4 \otimes \diag\left(\mathds{1}_{2N}, \ytens\right)\right) \diag\left(X \otimes\mathds{1}_{2N}, X\otimes\mathds{1}_{2N}, \mathds{1}_{8N}\right) \widetilde{\bm{V}}_s .
\end{align}
The inverse of $\Sigma$ directly follows from the derivation in Ref. \cite{GG2025} as no assumptions on the supermatrix $\sigma$ are made. Thus, we replace $\sigma$ by $\sigma(\xi)$ and have
\begin{align}
	\Sigma^{-1} =& \sigma_E(\xi)^{-1} \otimes \mathds{1}_{2N} - \sigma_E(\xi)^{-1} \otimes \sum_{c=1}^{M} \frac{\pi}{\gamma_c} W_c W_c^\dagger + \sum_{c=1}^{M} \rho^{(c)}(\xi) \otimes \frac{\pi}{\gamma_c} W_c W_c^\dagger, \notag\\
	\rho^{(c)}(\xi) =& \left(\sigma_E(\xi) + \frac{i\gamma_c}{8\pi k} L\right)^{-1} .
\end{align} 
This implies that the phase 
\begin{equation}
	F_s^{(4)}(\xi) = \widehat{\bm{V}}_s^\trans \left(\widetilde{L}^{-1/2} \otimes\mathds{1}_{2N}\right) \Sigma^{-1} \left(\widetilde{L}^{-1/2} \otimes\mathds{1}_{2N}\right) \overline{\bm{V}}_s
\end{equation}
is given in terms of elements of $\rho^{(c)}(\xi)$ due to the orthogonality of different channels
\begin{equation}
	F_s^{(4)}(\xi) = i^{s+1} \frac{\gamma_b}{\pi} \left(\rho^{(b)}_{13}(\xi) + \rho^{(b)}_{42}(\xi)\right) + (-i)^{s+1} \frac{\gamma_a}{\pi} \left(\rho^{(a)}_{31}(\xi) + \rho^{(a)}_{24}(\xi)\right) .
\end{equation}
Anticipating a saddle point approximation as our next step we choose to factorize the determinant in \cref{eqn:intRealSupervector}
\begin{align}
	\sdet \Sigma =& \sdet^{2N} \sigma_E \sdet^{2N} \left(\mathds{1}_8 + \frac{\pi^2 \xi^2}{N} \sigma_E^{-1} \fourotau \sigma \fourotau\right) \notag\\
	&\times\prod_{c=1}^{M} \sdet\left(\mathds{1}_8 + \frac{i\gamma_c}{8\pi k} \sigma_E(\xi)^{-1} L\right)
\end{align}
using the product property of the superdeterminant.
\subsection{Saddle Point Approximation}
\label{subsec:SaddlePoint}
The mapping on superspace makes it possible to carry out the limit $N\to\infty$ by means of a saddle point approximation. To that end we divide the integrand into two parts
\begin{equation}
	R_{s,m m\myprime}(k\vert \xi) = \mathcal{N} \int \dd[\sigma] \exp\left(- N \mathcal{L} - \delta \mathcal{L}\right).
\end{equation}
The dominant part determining the saddle point
\begin{equation}
	\mathcal{L} = \frac{(8\pi k)^2}{2v^2} \str\sigma^2 + \str\ln\sigma_E
\end{equation}
and the fluctuations are contained in
\begin{align}\label{eqn:GSEFluctuations}
	\delta\mathcal{L} =& \frac{\left(8\pi k\right)^2 \pi^2 \xi^2}{2v^2} \str\left(\fourotau \sigma\right)^2 + N \str \ln \left(\mathds{1}_8 + \frac{\pi^2\xi^2}{N} \sigma_E^{-1} \fourotau \sigma \fourotau\right) \notag\\
	&+ \sum_{c=1}^{M} \str\ln\left(\mathds{1}_8 + \frac{i \gamma_c}{8k} \sigma_E(\xi)^{-1} L\right) \notag\\
	&+ \frac{i}{16} \left(i^{s+1} \frac{\gamma_b}{\pi} \left(\rho_{13}^{(b)}(\xi) + \rho_{42}^{(b)}(\xi)\right) + (-i)^{s+1} \frac{\gamma_a}{\pi} \left(\rho_{31}^{(a)}(\xi) + \rho_{24}^{(a)}(\xi)\right)\right) .
\end{align}
While it might seem that the second term of the fluctuations linearly depends on $N$ and should belong to the dominant part, expansion of the logarithm in powers of the argument shows that this is not the case. Clearly, the saddle point equation is the same as the in the case of unbroken time reversal invariance. Hence, the solutions are also the same
\begin{equation}\label{eqn:diagonalSaddlePointSolution}
	\sigma_D^0 = \frac{E}{16\pi k} \mathds{1}_8 + \frac{i \Delta}{16\pi k} L, \quad \Delta = \sqrt{4v^2 - E^2}
\end{equation}
and the full saddle point manifold is encompassed by
\begin{equation}
	\sigma_G = T^{-1} \sigma_D^0 T = \frac{E}{16\pi k} \mathds{1}_8 - \frac{\Delta}{16\pi k} Q, \quad Q = -i T^{-1} L T.
\end{equation}
The integration decouples into integrations over transformations of the saddle point, ``Goldstone" modes, and deviations from the saddle point, ``massive" modes. The latter amount to a Gaussian integration in superspace yielding a numerical prefactor which we absorb into the normalization. We are left with the non-linear sigma model
\begin{align}\label{eqn:nonLinearSigma}
	R_{s}(k \vert \xi) =& \mathcal{N}\int\dd\mu(Q) \exp\left(+r \pi^2\xi^2 \str \left(\fourotau \sigma_G\right)^2\right)  \notag\\
	&\times \prod_{c=1}^{M}\sdet^{-1}\left(\mathds{1}_8 + \frac{i \gamma_c}{8k} \sigma_{G,E}^{-1} L\right) \exp\left(-\frac{i}{16} F_{s}^{(4)}\right) 
\end{align}
as integral over the coset space $\text{UOSp}(2,2\vert 4)/\text{UOSp}(2\vert 2) \times \text{UOSp}(2\vert 2)$.
At this point we have dropped the indices $m, m\myprime$ as the characteristic function does not depend on the spin orientation, and we will also drop the index $G$ as it is clear that all subsequent calculations are performed at the saddle point. Before proceeding, we point out that it is quite remarkable that compared to the case of unbroken time reversal invariance the non-linear sigma model for the transitional ensemble in \cref{eqn:nonLinearSigma} only contains an additional exponential factor. This was already observed for time reversal invariance breaking in systems with orthogonal symmetry in Ref. \cite{PWZLW1995} where the authors calculated correlations of scattering matrix elements. We also emphasize the sign change in front of the supertrace as compared to \cref{eqn:GSEFluctuations} which is a result of an additional contribution arising from the expanded second term.

\subsection{Integration over the Saddle Point Manifold}
\label{subsec:IntegrationSaddlePoint}
At the saddle point the invariance breaking term simplifies to
\begin{equation}\label{eqn:invBreakingSaddlePoint}
	\str\left(\fourotau \sigma\right)^2 = \left(\frac{\Delta}{16\pi k}\right)^2 \str \left(\fourotau Q\right)^2 .
\end{equation}
This means the full prefactor of the supertrace in \cref{eqn:nonLinearSigma} is 
\begin{equation}\label{eqn:Xi}
	\Xi = \pi^2\xi^2\Delta^2/(8v^2) .
\end{equation}
Furthermore, $\Delta/(2\pi v^2)$ is the eigenvalue density of the system at hand such that $\xi$ is measured on the scale of the local mean level spacing. As the other factors are the same as in the case of time reversal invariance we only have to determine the invariance breaking factor in terms of the parametrization in Ref. \cite{GG2025}, \textit{cf.} \ref{app:parametrization}. Inserting the definition of $Q$ into \cref{eqn:invBreakingSaddlePoint} and expanding the exponential in terms of the anticommuting variables is quite messy. Hence, we make use of a clever substitution introduced in Refs. \cite{AIMW1992,PWZLW1995} for systems with orthogonal symmetry. To that end we observe that the supertrace in \cref{eqn:nonLinearSigma,eqn:invBreakingSaddlePoint} is
\begin{align}
	-\str \left(\fourotau Q\right)^2 =& \str \left(Q_0 \mathcal{U} \fourotau \mathcal{U}^{-1}\right)^2 \notag\\
	=& \str\left(\cos\widehat{\theta} \mathcal{U}_1 \twootau \mathcal{U}_1^{-1} \right)^2 + \str\left(\cos\widehat{\theta} \mathcal{U}_2 \twootau \mathcal{U}_2^{-1} \right)^2 \notag\\
	&+ 2 \str \left(\sin\widehat{\theta} \mathcal{U}_1 \twootau \mathcal{U}_1^{-1}\right)\left(\sin\widehat{\theta} \mathcal{U}_2 \twootau \mathcal{U}_2^{-1}\right)
\end{align}
and notice that the anticommuting variables only appear in the form
\begin{equation}\label{eqn:vtauvinv}
	v_j \twootau v_j^{-1}.
\end{equation}
Introducing the unitary transformations $\exp(\eta_j)$ made up of the anticommuting variables $\mu_1, \nu_1$ introduced in \ref{app:parametrization},
\begin{equation}
	\eta_j = \begin{bNiceMatrix}
		0 & \mu_1^\star \nu_1^\star \\
		-\mu_1 \nu_1 & 0
	\end{bNiceMatrix} ,
\end{equation}
and the matrices
\begin{align}\label{eqn:gMat}
	g_1 =& \begin{bNiceMatrix}
		1+2\nu_1\nu_1^\star & 0 & 0 & 2 \nu_1^\star \\
		0 & -1-2\nu_1\nu_1^\star & 2\nu_1 & 0 \\
		0 & -2\nu_1^\star & 1- 2\nu_1\nu_1^\star & 0 \\
		2\nu_1 & 0 & 0 & -1 + 2\nu_1\nu_1^\star
	\end{bNiceMatrix} ,\notag\\
	g_2 =& \begin{bNiceMatrix}
		1-2\nu_2\nu_2^\star & 0 & 0 & 2 i \nu_2^\star \\
		0 & -1+2\nu_2\nu_2^\star & 2 i \nu_2 & 0 \\
		0 & -2 i \nu_2^\star & 1+ 2\nu_2\nu_2^\star & 0 \\
		2i\nu_2 & 0 & 0 & -1 - 2\nu_2\nu_2^\star
	\end{bNiceMatrix} ,
\end{align}
which exhibit the hyperbolic symmetry as $g_1 = g_1^\dagger$ and $g_2 = \left(\tauthree \otimes \mathds{1}_2\right) g_2^\dagger \left(\tauthree \otimes \mathds{1}_2\right)$. We find that \cref{eqn:vtauvinv} is given by a unitary rotation of the matrices $g_1, g_2$, respectively, 
\begin{align}
	v_1 \twootau v_1^{-1} =& \begin{bNiceMatrix}
		\left(e^{\eta_1}\right)^\dagger & 0 \\
		0 & \mathds{1}_2
	\end{bNiceMatrix} g_1 \begin{bNiceMatrix}
	e^{\eta_1} & 0 \\
	0 & \mathds{1}_2
	\end{bNiceMatrix} ,\notag\\
	v_2 \twootau v_2^{-1} =&  \begin{bNiceMatrix}
		e^{\eta_2} & 0 \\
		0 & \mathds{1}_2
	\end{bNiceMatrix} g_2  \begin{bNiceMatrix}
	\left(e^{\eta_2}\right)^\dagger & 0 \\
	0 & \mathds{1}_2
	\end{bNiceMatrix} .
\end{align}
Then we absorb the unitary transformations into the $\text{SU}(2)$ matrix $U$ by setting $U\myprime = \exp\left(\eta_1\right) U \exp\left(\eta_2\right)$. This greatly simplifies the expansion of the exponential as the matrices in \cref{eqn:gMat} only depend on four of the eight anticommuting variables and the Berizinian associated with the transformation is unity. However, we have to be careful as $F_s^{(4)}$ also depends on elements of $U$ such that we have to express them in terms of the new variables $U\myprime$. In the new variables the invariance breaking term amounts to 
\begin{align}\label{eqn:IBexplicit}
	&\str\left(\fourotau Q\right)^2 = \notag\\
	& - 4 \cosh^2\theta + 4 \cos\left(\theta_1 + \theta_2\right) \cos\left(\theta_1 - \theta_2\right) \notag\\
	&- 8 \left(\cosh^2\theta - 2 \cosh\theta \cos\theta_1 \cos\theta_2 + \cos\left(\theta_1 + \theta_2\right) \cos\left(\theta_1 - \theta_2\right)\right) \left(\nu_1 \nu_1^\star - \nu_2\nu_2^\star\right) \notag\\
	&+ 4 \sinh^2\theta \left(U_{11}\myprime U_{22}\myprime + U_{12}\myprime U_{21}\myprime\right) \left(1 + 2 \nu_1 \nu_1^\star\right)\left(1 - 2\nu_2 \nu_2^\star\right) \notag\\
	&+ 4 \sin\left(\theta_1 + \theta_2\right) \sin\left(\theta_1 - \theta_2\right) \left(1- 2 \nu_1 \nu_1^\star\right)\left(1 + 2 \nu_2 \nu_2^\star\right) \notag\\
	&- 16  \sinh\theta \cos\theta_2 \sin\theta_1 \left(U_{11}\myprime e^{i\left(\phi_1 - \phi_2\right)}\nu_1 \nu_2^\star - U_{22}\myprime e^{-i\left(\phi_1 - \phi_2\right)}\nu_1^\star \nu_2\right) \notag\\
	&- 16 \sinh\theta \cos\theta_1 \sin\theta_2 \left(U_{12}\myprime e^{i\left(\phi_1 + \phi_2\right)}\nu_1 \nu_2 - U_{21}\myprime e^{-i\left(\phi_1 + \phi_2\right)}\nu_1^\star \nu_2^\star\right) .
\end{align}
Using the results from Ref. \cite{GG2025}, we are able to expand in the anticommuting variables by taking $F_s^{(4)}$ and expressing the elements of $U$ by the elements of $U\myprime$. While the above transformation greatly simplifies the calculations they are still very lengthy and require precise bookkeeping. We use \textsc{Mathematica} \cite{Mathematica} to perform the expansion and some of the subsequent calculations. Furthermore, we refrain from showing intermediate results as they on the one hand do not carry any meaningful insights compared to the final expression and on the other hand bloat the present work with expressions stretching over multiple pages. Instead, we list the steps necessary to arrive at the final expression.

We start by expanding the integrand (\ref{eqn:nonLinearSigma}) in the anticommuting variables and then carry out the integration over them. This leaves only the terms depending on all eight anticommuting variables and all other terms vanish. Next, we do the trivial integral over the orthogonal angles $\phi_1, \phi_2$ as they do not appear under the integral. Similarly, the integration over the unitary angle $\varphi_2$ is also trivial as it appears nowhere. To carry out the integrals over $\varphi_1$ we use that for some non-negative integer $n$
\begin{equation}
	\int_{0}^{2\pi} \dd\varphi_1 \exp\left(\pm i n \varphi_1 + c_1 e^{i \varphi_1} + c_2 e^{-i \varphi_1}\right) = \frac{2\pi}{\sqrt{c_1 c_2}^n} I_n(2 \sqrt{c_1 c_2}) \begin{cases}
		c_2^n, & + \\
		c_1^n, & -
	\end{cases}
\end{equation}
where $I_n(z)$ is the modified Bessel function of first kind. This identity directly follows when expanding the exponential function in its power series, carrying out the integration over the resulting phases and identifying the resummation with the power series of aforementioned Bessel function. Unfortunately, it turns out that the integration over the radial unitary variable $u$ amounts to solving integrals of the form
\begin{equation}
	\int_0^1 \dd u \, u^n I_m(c_1 u) \exp\left(c_2 u^2\right) .
\end{equation}
To the best of our knowledge solutions to these integrals only exist involving recursive definitions, see for example \cite{Rosenheinrich2025}. Hence, we refrain from an exact integration and instead expand the integrand in orders of the parameter $\Xi=\pi^2\xi^2\Delta^2/(8v^2)$ as defined in \cref{eqn:Xi}. This approach is justified as in most experimentally relevant cases the breaking of the time reversal invariance is weak such that first order corrections are sufficient. Nevertheless, higher orders are easily calculated by expanding the exponential to desired order. The calculation of higher order terms are as difficult as those of the first order, they only involve a greater amount of terms. In this first order expansion the characteristic function splits into two terms 
\begin{equation}\label{eqn:cfFinal}
	R(k \vert \Xi) = R^{\text{GSE}}(k) + \Xi R^{\text{IB}}(k)
\end{equation}
where the first term is the result for system with time reversal invariance found in \cite{GDG2025,GG2025}, and the second one is the one breaking this invariance. We carry out the integration over $u$ which results in Bessel functions of varying orders. We reduce all order to zeroth and first order using recurrence relations. This leads us to our final expression for the invariance breaking term
\begin{align}\label{eqn:cfIB}
	R^{\text{IB}}(k) &= \int \dd[\widehat{\theta}] \frac{\sin \theta_1 \sin\theta_2 \sinh^3 \theta}{\left(\cos\left(\theta_1 + \theta_2\right) - \cosh\theta\right)^2 \left(\cos\left(\theta_1 - \theta_2\right) - \cosh\theta\right)^2} \notag\\
	&\times \prod_{c=1}^{M} \frac{\left(g_c^+ + \cos\left(\theta_1 + \theta_2\right)\right) \left(g_c^+ + \cos\left(\theta_1 - \theta_2\right)\right)}{\left(g_c^+ + \cosh\theta\right)^2} \notag\\
	&\times \left(\iota_0 J_0\left(\omega_{ab} k \right) + \iota_1 \frac{J_1\left(\omega_{ab} k \right)}{\omega_{ab} k}\right) .
\end{align}
The coefficients are shown in \ref{app:coefficientsGSE}. We discuss a few observations. The characteristic function, as in the case of unbroken time reversal invariance, does not depend on the spin orientations $m, m\myprime$. Intuitively, it is clear that this should not change when breaking time reversal invariance as systems without time reversal invariance do not distinguish between different spin orientations. Additionally, the characteristic functions for real and imaginary part are the same, this is why we dropped the index $s$ above. As both the symplectic $(\xi=0)$ and the unitary limit $(\xi\to\infty)$ exhibit this behavior, \textit{cf.} \cite{KNSGDMRS2013,GDG2025}, we anticipated this result. Furthermore, we spot that the integrand is invariant under the exchange of the two channels $a$ and $b$ as is also the case for intact and broken time reversal invariance. While the independence of spin orientation, equal distribution of real and imaginary part and invariance under exchange of $a, b$ are expected results, it is nevertheless satisfying that our calculations explicitly reproduce them. 

As obvious from \cref{eqn:cfIB} the invariance breaking term does not comprise additional Efetov-Wegner terms because the supertrace in \cref{eqn:nonLinearSigma} vanishes at the boundaries of integration. Hence, only the Efetov-Wegner term already present for conserved time reversal invariance appears \cite{GG2025}.

Furthermore, we do not explicitly show that $\xi\to\infty$ yields the unitary limit. Instead, we sketch the steps necessary to explicitly carry out the limit. Prior to the saddle point approximation the limit is rather trivial as it amounts to $\pi^2\xi^2/N\to1$ which effectively reduces the dimension of $\sigma(\xi)$ by a factor two and removes all additional symmetries besides $\sigma^\dagger = \widetilde{L} \sigma \widetilde{L}$. This yields the characteristic function for GUE systems, \textit{cf.} Ref. \cite{NKSG2014}. After carrying out the saddle point approximation the limit fixes the variables $u$ and $\theta_2$ and makes their integration trivial. Similar to Ref. \cite{PWZLW1995} we see that the non-linear sigma model of \cref{eqn:nonLinearSigma} has non-zero contributions only if the commuting part of \cref{eqn:IBexplicit} vanishes. Effectively, this means $u\sim 1$ and $\sin^2\theta_2\sim 0$, and we expand the rest of the integrand accordingly by only keeping the lowest orders. Performing these calculations will then result in the expression of Ref. \cite{KNSGDMRS2013}.

We also mention some further results that follow directly from our characteristic function. This is first and foremost the distribution which we obtain as
\begin{equation}
	P(x) = \frac{1}{2\pi} \int_{-\infty}^{\infty}\limits \dd k R(k) \exp\left(i k x\right) .
\end{equation}
The moments of the distribution are given by the derivatives of the characteristic function in \cref{eqn:cfFinal} at $k=0$ up to some prefactor. 
\subsection{Distribution of Scattering Cross Sections and Correlators}
\label{subsec:CrossSectionsGSE}
Importantly, the distribution of scattering cross sections is readily accessible from our results in the same fashion as in Ref. \cite{GG2025}. It is given as
\begin{equation}\label{eqn:besseltransformcross}
	p(\sigma_{am bm\myprime}\vert\Xi) = \frac{1}{4\pi} \int\dd^2\bm{k} J_0\left(\sqrt{\sigma_{am bm\myprime}} \lvert \bm{k} \rvert\right) R(\bm{k}\vert\Xi) ,
\end{equation}
the Bessel transform of the bivariate characteristic function $R(\bm{k})$. The bivariate characteristic function in the present case is a function of the absolute value $\vert\bm{k}\vert$ only, independent of $m,m\myprime$. Replacing $k$ by $\vert\bm{k}\vert$ in \cref{eqn:cfFinal} gives $R(\bm{k}\vert \Xi)$. As already alluded to, it is also in principle possible to determine higher orders in $\xi$ without any further technical complications. The only constraint is the ability to handle the large amount of terms in a meaningful manner.

We mention that the correlations of scattering matrix elements for this model were calculated in Ref. \cite{PWZLW1995} for time reversal invariance breaking in system with orthogonal symmetry. All the technically challenging aspects of such calculations were already performed in the present work and Ref. \cite{GG2025}. Additionally, unlike the characteristic functions the calculation of the correlations does not involve expanding an exponential of anticommuting variables. Thus, some of the difficulties present for the characteristic function are not present for the correlations. However, the calculation of such correlations is part of ongoing research and goes beyond the scope of the present work.

\section{Time Reversal Invariance Breaking for Orthogonal Symmetry}
\label{sec:goegueScatterMat}
Following a similar structure as in the previous section, we begin by defining the distribution of the off-diagonal scattering matrix elements in \cref{subsec:DistributionGOE}. In \cref{subsec:HubStrTransformation} we map the characteristic function on superspace. Subsequently, we carry out a saddle point approximation in \cref{subsec:SaddlePointGOE} and integrate over the resulting saddle point manifold in \cref{subsec:IntegrationSaddlePointGOE} together with a discussion of the resulting expressions. Concluding this section, we show an expression for the scattering cross sections in \cref{subsec:ScatteringCrossSectionsGOE}. 
\subsection{Distribution of  Off-Diagonal Scattering Matrix Elements}
\label{subsec:DistributionGOE}
Systems with orthogonal symmetry are of greater experimental relevance compared to systems with symplectic symmetry. Hence, we want to calculate the distributions of off-diagonal scattering matrix elements for time reversal invariance breaking in orthogonal systems as well. Unlike in the symplectic case a great amount of work was already done in Ref. \cite{PWZLW1995} where the authors calculated the correlations of scattering matrix elements. Many steps directly carry over to the characteristic function. Additionally, some of the above calculations for symplectic systems as well as those for intact time reversal invariance from Ref. \cite{NKSG2014} are equally relevant. Thus, we elect to give only a small amount of detail as to not repeat large parts of Refs. \cite{PWZLW1995,NKSG2014}. 

We setup the time reversal invariance breaking according to \cref{eqn:goeIB} such that in this case we only have to determine two distributions defined as
\begin{equation}
	P_s(x_s\vert \xi) = \int\dd[H] \mathcal{P}(H) \delta\left(x_s - \wp_s\left(S_{ab}(E)\right)\right)
\end{equation}
where $\wp_s\left(S_{ab}(E)\right)$ is the real, imaginary part of the scattering matrix elements for $s=1,2$, respectively. Here, the distribution $\mathcal{P}(H)$ is
\begin{align}
	\mathcal{P}(H) =& \int\dd[H_{\text{GOE}}] \int\dd[H_{\text{skewGOE}}] \delta\left(H - \left(H_{\text{GOE}} + i \frac{\pi \xi}{\sqrt{N}} H_{\text{skewGOE}}\right)\right) \notag\\
	&\times \mathcal{P}^{(1)}\left(H_{\text{skewGOE}}\right) \mathcal{P}^{(1)}_{\text{skew}}\left(H_{\text{skewGOE}}\right), \notag\\
	\mathcal{P}^{(1)}_{\text{skew}}\left(H_{\text{skewGOE}}\right) \sim& \exp\left(+ \frac{N}{4v^2} \tr H_{\text{skewGOE}}^2\right)
\end{align}
with a positive sign for the distribution of the skew-symmetric matrix $H_{\text{skewGOE}}$ as its square is negative semi-definite. A Fourier transform results in
\begin{equation}
	R_s(k\vert \xi) = \int\dd[H] \mathcal{P}(H) \exp\left(-ik \wp_s\left(S_{ab}(E)\right)\right),
\end{equation}
the corresponding characteristic function.

\subsection{Hubbard-Stratonovich Transformation}
\label{subsec:HubStrTransformation}
Replacing the phase by integrals over $2N$-dimensional commuting vectors $z=(z_a, z_b)$ and anticommuting vectors $\zeta=(\zeta_a, \zeta_b)$ and carrying out a Hubbard-Stratonovich transformation after the fact works exactly as shown in Refs. \cite{NKSG2014}. However, unlike the authors of the latter we choose to work with complex instead of real variables. This means we obtain for the characteristic function
\begin{align}
	R_s(k\vert \xi) =& (-1)^N\mathcal{N} \int\dd[\sigma] \exp\left(- N \widetilde{r} \str \left\{\sigma^2 + \frac{\pi^2 \xi^2}{N} \left(\fourotau \sigma\right)^2\right\}\right) \notag\\
	&\times \int\dd[\Psi] \exp\left(i \Psi^\dagger \left(\widetilde{L}^{1/2} \otimes \mathds{1}_N\right) \Sigma \left(\widetilde{L}^{1/2} \otimes \mathds{1}_N\right) \Psi + \frac{i}{2} \Psi^\dagger \bm{V}_s\right), \notag\\
	\Psi =& \left(z_a, z_a^\star, z_b, z_b^\star, \zeta_a, \zeta_a^\star, \zeta_b, \zeta_b^\star\right), \notag\\
	\Sigma =& \sigma_E(\xi) \otimes \mathds{1}_N + \frac{i}{8k} L \otimes \sum_{c=1}^{M} W_c W_c^\dagger, \notag\\
	\bm{V}_s =& \left(V_s, 0\right), \quad V_s=\left(W_a, -i^s W_b, W_b, (-i)^s W_a\right)
\end{align}
and $\widetilde{r}=(4\pi k)^2/v^2$. Due to our choice of sticking with complex instead of real coordinates the symmetries of $\sigma$ are also altered. This means to ensure $\sigma$ inherits the correct symmetries from $B$ we have
\begin{align}
	\sigma =& T^{-1} \mathfrak{V} \diag\left(\sigma_{\text{F},1},\sigma_{\text{F},2},\sigma_{\text{F},3},\sigma_{\text{F},4},i\sigma_{\text{B},1}\mathds{1}_2, i\sigma_{\text{B},2}\mathds{1}_2\right) \mathfrak{V}^\dagger T, \notag\\
	T^\dagger \widetilde{L} T =& \widetilde{L}, \quad C T^\star C^\trans = T, \quad C = \diag\left(X, -X, Y, Y\right), \notag\\
	X=&\tau^{(1)}, \quad \mathfrak{V}= \diag\left(\mathds{1}_2\otimes\mathfrak{v},\mathds{1}_4\right), \quad \mathfrak{v}=\frac{1}{\sqrt{2}} \begin{bNiceMatrix}
		1 & i \\
		1 & -i
	\end{bNiceMatrix} .
\end{align}
Notice here that the bosonic and fermionic subspaces for the diagonal elements of $\sigma$ and the blocks of $C$ and $\mathfrak{V}$ are swapped when compared to \cref{eqn:gseSigmaDiag}. This was extensively discussed in Ref. \cite{GG2025} where we found that systems with orthogonal and symplectic symmetry are closely connected and in superspace this is reflected by switching of the roles of fermionic and bosonic subspaces, as long as similar coordinates are chosen. On the same note the transformations $T$ again form the non-compact orthosymplectic supergroup $\text{UOSp}(2,2\vert 4)$. The integration over the supervector $\Psi$ is simple as our $\Sigma$ is related to the $\Sigma$ in Ref. \cite{NKSG2014} by the transformation $\mathfrak{V}$ for $\xi=0$. Straightforward calculations show that the additional invariance breaking term also fulfills the same symmetries and the integration yields
\begin{align}
	&\int\dd[\Psi] \exp\left(i \Psi^\dagger \left(\widetilde{L}^{1/2} \otimes \mathds{1}_N\right) \Sigma \left(\widetilde{L}^{1/2} \otimes \mathds{1}_N\right) \Psi + \frac{i}{2} \Psi^\dagger \bm{V}_s\right) \notag\\
	=& (-1)^N \sdet^{-1/2}\Sigma \exp\left(-\frac{i}{16} F_s^{(1)}(\xi)\right) 
\end{align}
with the phase
\begin{equation}\label{eqn:GOEPhase}
	F_s^{(1)}(\xi)= \bm{V}_s^\trans \left(\diag\left(X,X,\mathds{1}_4\right)\otimes\mathds{1}_N\right) \left(\widetilde{L}^{-1/2} \otimes \mathds{1}_N\right) \Sigma^{-1} \left(\widetilde{L}^{-1/2} \otimes \mathds{1}_N\right) \bm{V}_s .
\end{equation}
Using the result for the inverse $\Sigma^{-1}$ from Ref. \cite{NKSG2014} we find that 
\begin{align}
	F_s^{(1)}(\xi)=& \frac{\gamma_a}{\pi} \left(\rho_{21}^{(a)}(\xi) + (-i)^{s+1} \rho_{24}^{(a)}(\xi) + (-i)^{s+1} \rho_{31}^{(a)}(\xi) - (-1)^s \rho_{34}^{(a)}(\xi)\right) \notag\\
	&+ \frac{\gamma_b}{\pi} \left((-1)^s\rho_{12}^{(b)}(\xi) + i^{s+1} \rho_{42}^{(b)}(\xi) + i^{s+1} \rho_{13}^{(b)}(\xi) - \rho_{43}^{(b)}(\xi)\right), \notag\\
	\rho^{(c)}(\xi) =& \left(\sigma_E(\xi) + \frac{i\gamma_c}{8\pi k} L\right)^{-1}.
\end{align}
Comparing with the phase found in Ref. \cite{NKSG2014} for $\xi=0$ we have a factor four less terms. This happens due to choosing complex coordinates which effectively leaves $\bm{V}_s$ as a real vector whereas for real coordinates $\bm{V}_s$ is rotated into the complex plane by the transformation $\mathfrak{V}$. As $\bm{V}_s$ appears twice in \cref{eqn:GOEPhase} it quadruples the amount of elements of $\rho^{(c)}$. Importantly, while it reduces the amount of terms we still obtain contributions from the off-diagonal and diagonal blocks of the Boson-Boson block contrary to the cases of symplectic and unitary symmetry where only the off-diagonal blocks contribute.

\subsection{Saddle Point Approximation and Parametrization of the Saddle Point Manifold}
\label{subsec:SaddlePointGOE}
We carry out a saddle point approximation where the same saddle point
\begin{equation}
	\sigma_G = \frac{E}{16\pi k} \mathds{1}_8 - \frac{\Delta}{16\pi k} Q, \quad Q = -i T^{-1} L T
\end{equation}
as in \cref{eqn:diagonalSaddlePointSolution} is relevant. Expanding the integrand around the saddle point and carrying out the integration over the massive modes leaves us with the non-linear sigma model
\begin{align}\label{eqn:nonLinearSigmaGOE}
	R_s(k \vert \xi) =& \mathcal{N} \int\dd\mu(Q) \exp\left(+ \Xi \str \left(\fourotau Q\right)^2\right)\notag\\
	&\times \prod_{c=1}^{M} \sdet^{-1/2}\left(\mathds{1}_8 + \frac{i\gamma_c}{8\pi k} \sigma_{G,E}^{-1} L\right) \exp\left(-\frac{i}{16} F_s^{(1)}\right) .
\end{align}
We are left with the integration over the saddle point manifold. To that end it is necessary to explicitly parametrize the matrix $Q$ which we show in \ref{app:parametrization}. Similar to the above case of symplectic symmetry and Refs. \cite{AIMW1992,PWZLW1995} the anticommuting variables in the invariance breaking factor from \cref{eqn:nonLinearSigmaGOE} only appear in the form of \cref{eqn:vtauvinv}. Hence, we introduce 
\begin{equation}
	\eta_j = \begin{bNiceMatrix}
		0 & -\mu_j \nu_j \\
		\mu_j^\star \nu_j^\star & 0
	\end{bNiceMatrix}
\end{equation}
and
\begin{align}
	g_1 =& \begin{bNiceMatrix}
		1 + 2 \nu_1 \nu_1^\star & 0 & 0 & -2\nu_1 \\
		0 & -1 -2 \nu_1 \nu_1^\star & -2\nu_1^\star & 0 \\
		0 & -2\nu_1 & 1-2\nu_1 \nu_1^\star & 0 \\
		2\nu_1^\star & 0 & 0 & -1 + 2 \nu_1 \nu_1^\star
	\end{bNiceMatrix} \notag\\
	g_1 =& \begin{bNiceMatrix}
		1 - 2 \nu_2 \nu_2^\star & 0 & 0 & -i2\nu_2 \\
		0 & -1 +2 \nu_2 \nu_2^\star & -2i\nu_2^\star & 0 \\
		0 & -2i\nu_1 & 1+2\nu_2 \nu_2^\star & 0 \\
		2i\nu_2^\star & 0 & 0 & -1 - 2 \nu_2 \nu_2^\star
	\end{bNiceMatrix}
\end{align}
such that
\begin{align}
	v_1 \twootau v_1^{-1} =& \begin{bNiceMatrix}
		\mathds{1}_2 & 0 \\
		0 & \left(e^{\eta_1}\right)^\dagger 
	\end{bNiceMatrix} g_1 \begin{bNiceMatrix}
	\mathds{1}_2 & 0 \\
	0 & e^{\eta_1}
	\end{bNiceMatrix} \notag\\
	v_2 \twootau v_2^{-1} =& \begin{bNiceMatrix}
		\mathds{1}_2 & 0 \\
		0 & e^{\eta_2}
	\end{bNiceMatrix} g_2 \begin{bNiceMatrix}
	\mathds{1}_2 & 0 \\
	0 & \left(e^{\eta_2}\right)^\dagger 
	\end{bNiceMatrix} .
\end{align}
Here, the matrices $g_1, g_2$ also reflect the hyperbolic symmetry as $g_1 = g_1^\dagger$ and $g_2 = \left(\tauthree \otimes \mathds{1}_2\right) g_2^\dagger \left(\tauthree \otimes \mathds{1}_2\right)$. Introducing new variables $U\myprime = \exp(\eta_1) U \exp(\eta_2)$ the invariance breaking term is
\begin{align}
	&\str \left(\fourotau Q\right)^2 \notag\\
	=& 4 \cos^2\theta - 4 \cosh\left(\theta_1 + \theta_2\right)\cosh\left(\theta_1 - \theta_2\right) \notag\\
	&- 8 \left( \cos^2\theta + \cosh\left(\theta_1 + \theta_2\right) \cosh\left(\theta_1 - \theta_2\right) - 2 \cos\theta \cosh\theta_1 \cosh\theta_2\right) \left(\nu_1 \nu_1^\star - \nu_2 \nu_2^\star\right) \notag\\
	&+ 4 \sin^2\theta \left(U_{11}\myprime U_{22}\myprime + U_{12}\myprime U_{21}\myprime\right) \left(1 - 2 \nu_1 \nu_1^\star\right)\left(1 + 2 \nu_2 \nu_2^\star\right) \notag\\
	&+ 4 \sinh\left(\theta_1 + \theta_2\right) \sinh\left(\theta_1 - \theta_2\right) \left(1 + \nu_1 \nu_1^\star\right)\left(1 - \nu_2 \nu_2^\star\right) \notag\\
	&+ 16 \sin\theta \cosh\theta_2 \sinh\theta_1 \left(U_{11}\myprime e^{i\left(\phi_1 - \phi_2\right)} \nu_1^\star\nu_2 - U_{22}\myprime e^{-i\left(\phi_1 - \phi_2\right)} \nu_1 \nu_2^\star \right) \notag\\
	&- 16 \sin\theta \cosh\theta_1 \sinh\theta_2 \left( U_{12}\myprime e^{i\left(\phi_1 + \phi_2\right)} \nu_1^\star \nu_2^\star - U_{21}\myprime e^{-i\left(\phi_1 + \phi_2\right)} \nu_1 \nu_2\right)
\end{align}
which is structurally very similar to the one for symplectic systems. Furthermore, it is also necessary to determine the phase $F_s^{(1)}$ in the chosen parametrization while also minding our change from $U$ to $U\myprime$. The phase is
\begin{align}
	- \frac{1}{4k}F_s^{(1)} =&q_a^+ \mu_1 \nu_1^\star + \frac{1}{2}e^{-2i\phi_1} q_a^- \left(1 + \mu_1 \mu_1^\star + \nu_1 \nu_1^\star - \mu_1 \mu_1^\star \nu_1 \nu_1^\star\right) \notag\\
	&-(-i)^{s+1}\left(1+ \frac{1}{2}\left(\mu_1 \mu_1^\star + \nu_1 \nu_1^\star\right) - \frac{3}{4} \mu_1 \mu_1^\star \nu_1 \nu_1^\star\right) \notag\\
	&\times\left(1- \frac{1}{2}\left(\mu_2 \mu_2^\star + \nu_2 \nu_2^\star\right) - \frac{3}{4} \mu_2 \mu_2^\star \nu_2 \nu_2^\star\right) \notag\\
	&\times \left(e^{i\left(\phi_1 - \phi_2\right)} p_a^+ \mu_1 \nu_1^\star \mu_2^\star \nu_2 + e^{-i\left(\phi_1 - \phi_2\right)} p_a^+ + e^{i\left(\phi_1 + \phi_2\right)} p_a^- \mu_1 \nu_1^\star + e^{-i\left(\phi_1 + \phi_2\right)} p_a^- \mu_2^\star \nu_2\right) \notag\\
	&- 2 (-i)^{s+1} p_a^0 \left(1 + \frac{1}{2} \left(\mu_1 \mu_1^\star - 3 \nu_1 \nu_1^\star\right) + \frac{3}{4} \mu_1 \mu_1^\star \nu_1 \nu_1^\star\right) \notag\\
	&\times \left(1 - \frac{1}{2} \left(\mu_2 \mu_2^\star - 3 \nu_2 \nu_2^\star\right) + \frac{3}{4} \mu_2 \mu_2^\star \nu_2 \nu_2^\star\right)\notag\\
	&\times\left(U_{11}\myprime \nu_1^\star \nu_2 - U_{22}\myprime \mu_1 \mu_2^\star + U_{12}\myprime \nu_1^\star \mu_2^\star - U_{21}\myprime \mu_1 \nu_2\right) \notag\\
	&- (-1)^s \left( r_a^+ \mu_2^\star \nu_2 +\frac{1}{2} e^{2i\phi_2} r_a^- \left(1 - \mu_2 \mu_2^\star - \nu_2 \nu_2^\star - \mu_2 \mu_2^\star \nu_2 \nu_2^\star\right)\right) \notag\\
	&+ (-1)^s \left(- q_b^+ \mu_1^\star \nu_1 + \frac{1}{2} e^{2i\phi_1} q_b^- \left(1 + \mu_1 \mu_1^\star + \nu_1 \nu_1^\star - \mu_1 \mu_1^\star \nu_1 \nu_1^\star\right)\right)\notag\\
	&- i^{s+1} \left(1 + \frac{1}{2} \left(\mu_1 \mu_1^\star + \nu_1 \nu_1^\star\right) - \frac{3}{4} \mu_1 \mu_1^\star \nu_1 \nu_1^\star\right) \notag\\
	&\times \left(1 - \frac{1}{2} \left(\mu_2 \mu_2^\star + \nu_2 \nu_2^\star\right) - \frac{3}{4} \mu_2 \mu_2^\star \nu_2 \nu_2^\star\right)\notag\\
	&\times \left(e^{i\left(\phi_1 - \phi_2\right)} p_b^+ + e^{-i\left(\phi_1 - \phi_2\right)} p_b^+ \mu_1^\star \nu_1 \mu_2 \nu_2^\star - e^{i\left(\phi_1 + \phi_2\right)} p_b^- \mu_2 \nu_2^\star - e^{-i\left(\phi_1 + \phi_2\right)} p_b^- \mu_1^\star \nu_1\right)\notag\\
	&- 2 i^{s+1} p_b^0 \left(1 + \frac{1}{2} \left(\mu_1 \mu_1^\star - 3 \nu_1 \nu_1^\star\right) + \frac{3}{4} \mu_1 \mu_1^\star \nu_1 \nu_1^\star\right) \notag\\
	&\times \left(1 - \frac{1}{2} \left(\mu_2 \mu_2^\star - 3 \nu_2 \nu_2^\star\right) + \frac{3}{4} \mu_2 \mu_2^\star \nu_2 \nu_2^\star\right)\notag\\
	&\times \left(U_{11}\myprime \mu_1^\star \mu_2 - U_{22}\myprime \nu_1 \nu_2^\star + U_{12}\myprime \mu_1^\star \nu_2^\star - U_{21}\myprime \nu_1 \mu_2\right)\notag\\
	&+ r_b^+ \mu_2 \nu_2^\star - e^{-2i\phi_2} r_b^- \left(1- \mu_2 \mu_2^\star - \nu_2 \nu_2^\star - \mu_2 \mu_2^\star \nu_2 \nu_2^\star\right)
\end{align}
where we use the same notation as Ref. \cite{NKSG2014}
\begin{align}\label{eqn:qprGOE}
	q_c^+ =& \left(\frac{E}{\Delta} + i g_c^-\right) \left(\frac{1}{g_c + \cosh\left(\theta_1 + \theta_2\right)} + \frac{1}{g_c + \cosh\left(\theta_1 - \theta_2\right)} - \frac{2}{g_c + \cos\theta}\right), \notag\\
	q_c^- =& \left(\frac{E}{\Delta} + i g_c^-\right) \left(\frac{1}{g_c + \cosh\left(\theta_1 + \theta_2\right)} - \frac{1}{g_c + \cosh\left(\theta_1 - \theta_2\right)}\right), \notag\\
	p_c^\pm =& \frac{\sinh\left(\theta_1 + \theta_2\right)}{g_c + \cosh\left(\theta_1 + \theta_2\right)} \pm \frac{\sinh\left(\theta_1 - \theta_2\right)}{g_c + \cosh\left(\theta_1 - \theta_2\right)}, \notag\\
	r_c^\pm =& \left(q_c^\pm\right)^\star, \quad p_c^0 = \frac{\sin\theta}{g_c + \cosh\theta}, \quad g_c^- = \frac{\gamma_c^2 - v^2}{\Delta \gamma_c} .
\end{align}
The next step is to expand both exponentials in \cref{eqn:nonLinearSigmaGOE} with respect to the anticommuting variables and then carry our as many of the integrals as possible. These calculations are extremely lengthy and again opt to use \textsc{Mathematica} \cite{Mathematica} to carry out the expansion and subsequent integration. As in the symplectic case we list our steps and present additional information necessary for the calculations.

\subsection{Integration over the Saddle Point Manifold}
\label{subsec:IntegrationSaddlePointGOE}
We begin with expanding the integrand in terms of the anticommuting variables and carrying out their integration. This yields only terms that contain all eight anticommuting variables and the integrals of all others vanish. As none of these terms depend on $\varphi_1, \varphi_2$ their integration is trivial. Interestingly, in this case we can also explicitly carry out the integration over $u$ as it appears as
\begin{equation}
	\mathfrak{u}_{2n+1} = \int_{0}^{1} \dd u \, u^{2n+1} \exp\left(8 \Xi \sin^2\theta u^2\right), \quad n=0,1,2,3 .
\end{equation}
Hence, in the orthogonal case it is not necessary to expand in $\Xi$, and we derive the characteristic function for arbitrary values of $\Xi$. This is because the commuting part of the phase $F_s^{(1)}$ is independent of the unitary degrees of freedom which makes the integration above trivial. However, the commuting part of $F_s^{(1)}$ does now depend on the orthogonal variables $\phi_1, \phi_2$. We follow Ref. \cite{NKSG2014} and introduce center and difference variables
\begin{equation}
	\Phi= \frac{1}{2} \left(\phi_1 + \phi_2\right), \quad \psi = \phi_1 - \phi_2, \quad \Phi, \psi \in[0,2\pi]
\end{equation}
where both variables range over the full $2\pi$ interval due to the $2\pi$-periodicity of the integration. In these variables the integral over $\psi$ is of the form
\begin{equation}
	\int_{0}^{2\pi} \dd\psi \exp\left(\pm i n \psi + a_1 e^{i\psi} + a_2 e^{-i\psi}\right) = 2\pi \frac{I_n(2\sqrt{a_1 a_2})}{\sqrt{a_1 a_2}^n} \begin{cases}
		a_2^n, & + \\
		a_1^n, & -
	\end{cases}
\end{equation}
which is solved by expanding the exponential in its power series before carrying out the integration and resumming the resulting terms. Here, we have $a_1 = k \mathcal{Y}$ and $a_2 = -k \mathcal{X}$ where we define in analogy to Ref. \cite{NKSG2014}
\begin{align}
	\mathcal{X} =& \frac{1}{8} \left(2 (-i)^{s} p_{a}^{+} - i e^{-2i \Phi} q_{a}^{-} + i (-1)^{s} e^{2i \Phi} r_{a}^{-} \right) , \notag\\
	\mathcal{Y} =& \frac{1}{8} \left( 2 i^{s} p_{b}^{+} - i e^{-2i \Phi} r_{b}^{-} + i (-1)^{s} e^{2i \Phi} q_{b}^{-} \right) .
\end{align}
Unfortunately, to the best of our knowledge the integration over $\Phi$ is not possible in a closed form. We briefly mention that technically both integrals are solvable by expanding the original expression as a power series leaving us with trivial phase integrations. However, the resummation does not yield any meaningful results but rather messy accumulation of different sums and complicated dependences of the summation indices on each other. Finally, we obtain the characteristic function as a four-fold integral
\begin{align}
	R_s(k \vert \xi) =& 1 + \frac{1}{\pi} \int_{0}^{\pi}\dd\theta \int_{0}^{\infty}\dd\theta_1 \int_{0}^{\infty}\dd\theta_2 \int_{0}^{2\pi} \dd\Phi \frac{\sinh\theta_1 \sinh\theta_2 \sin^3\theta}{\left(\cosh\left(\theta_1 + \theta_2\right)-\cos\theta\right)^2 \left(\cosh\left(\theta_1 - \theta_2\right)-\cos\theta\right)^2} \notag\\
	&\times \prod_{c=1}^{M} \frac{g_c + \cos\theta}{\sqrt{g_c + \cosh\left(\theta_1 + \theta_2\right)}\sqrt{g_c + \cosh\left(\theta_1 - \theta_2\right)}} \notag\\
	&\times \exp\left(-2 \Xi\left(\sinh^2\theta_2 + \sin^2\theta \right)\right) \sum_{j=0}^{4} \bm{\iota}_j I_j(i \omega k)
\end{align}
where $\omega=2\sqrt{\mathcal{X}\mathcal{Y}}$. The coefficients $\bm{\iota}_j$ are very lengthy we thus relegate them to \ref{app:coefficients}. As in the case of unbroken time reversal invariance and the transition in GSE systems there exist a symmetry $a\leftrightarrow b$. Performing the substitution $\Phi \to -\Phi + \pi/2$ shows that the terms are symmetric due to the form of the coefficients $\bm{\iota}_j$. This is at first glance somewhat surprising as experimental measurement in GOE systems with weakly broken time reversal invariance showed that $S_{ab}$ and $S_{ba}$ are not equal \cite{DFH2009}. Importantly, this does not stand in conflict with our result because we obtain that the distribution is symmetric under the exchange $a\leftrightarrow b$ but that does not imply that in any specific realization of the scattering matrix the elements $S_{ab} = S_{ba}$ are symmetric. Indeed, this is a very interesting result which shows that the reciprocity of the distribution is a much weaker condition than the reciprocity of the scattering matrix. Thus, it is not possible to analyse time reversal invariance breaking through comparison of the distributions of $S_{ab}$ and $S_{ba}$. Instead, our results provide a considerably easier access to investigate time reversal invariance breaking by analysing the distribution of one scattering matrix element $S_{ab}$. In particular the moments are easily extracted from experimental data and allow us to go beyond linear correlations, as alluded to in \cref{sec:timeRevInvBreaking}. As this point is particularly important we want to emphasize that in deriving the characteristic function we have made it possible to analyse the strength of time reversal invariance breaking without the need for detailed balance experiments. Instead, it is sufficient to extract the statistics of a single scattering matrix elements and detailed balance analysis can be used as auxiliary data to confirm that the system is well described by RMT. Furthermore, the integral contains only even orders in $k$ and is therefore symmetric under $k\to -k$. Importantly, the integral clearly depends on $s$ contrary to the GSE case, which shows that for finite breaking of the time reversal invariance real and imaginary parts of the scattering matrix elements are not equally distributed. This also explains the appearance of $E/\Delta + i g_c^-$ in \cref{eqn:qprGOE} as they are related to the phase of the average diagonal elements $\overline{S}_{cc}$. Hence, due to the different distributions of real and imaginary part additional phase information is necessary which is not the case for unitary and symplectic system as already argued in Ref. \cite{NKSG2014}. Here, this is not changed either by breaking the time reversal invariance.

As a last point, we mention that we recover the GUE case in the limit $\Xi\to\infty$ both before the saddle point approximation and after. Similar to the GSE case the limit $\Xi\to\infty$ fixes the variables $\cosh\theta_2$ and $u$ to the proximity of '1'. Hence, we have to expand the integral accordingly and only keep the highest order in $\Xi$. The resulting expression is exactly the one obtained in \cite{NKSG2014} for GUE systems. Furthermore, real and imaginary part become equally distributed as expected, and we conclude that highest order corrections are independent of $s$. Hence, this means there must be some threshold invariance breaking strength $\Xi_{\text{T}}$ above which real and imaginary parts become effectively equally distributed.

\subsection{Distribution of Scattering Cross Sections}
\label{subsec:ScatteringCrossSectionsGOE}

We are also able to calculate the distribution of the cross sections in a similar fashion to \cite{NKSG2014,GG2025} and obtain a similar result for the bivariate characteristic function $R_s(\bm{k}\vert \xi)$ as in the univariate case
\begin{align}
	R(\bm{k} \vert \xi) =& 1 + \frac{1}{\pi} \int_{0}^{\pi}\dd\theta \int_{0}^{\infty}\dd\theta_1 \int_{0}^{\infty}\dd\theta_2 \int_{0}^{2\pi} \dd\Phi \frac{\sinh\theta_1 \sinh\theta_2 \sin^3\theta}{\left(\cosh\left(\theta_1 + \theta_2\right)-\cos\theta\right)^2 \left(\cosh\left(\theta_1 - \theta_2\right)-\cos\theta\right)^2} \notag\\
	&\times \prod_{c=1}^{M} \frac{g_c + \cos\theta}{\sqrt{g_c + \cosh\left(\theta_1 + \theta_2\right)}\sqrt{g_c + \cosh\left(\theta_1 - \theta_2\right)}} \notag\\
	&\times \exp\left(-2 \Xi\left(\sinh^2\theta_2 + \sin^2\theta \right)\right) \sum_{j=0}^{4} \bm{\iota}_j I_j(i \omega(\bm{k})) .
	\end{align} 
Here, we have $\omega(\bm{k}) = \sqrt{\mathcal{X}(\bm{k}) \mathcal{Y}(\bm{k})}$ with
\begin{align}\label{eqn:XYCrossSections}
	\mathcal{X}(\bm{k}) =& \frac{1}{8} \left(-2 i \absk p_a^+ - e^{-2i\Phi} \bm{k} q_a^- + e^{2i\Phi} \bm{k}^\star r_a^-\right) \notag\\
	\mathcal{Y}(\bm{k}) =& \frac{1}{8} \left(2 i \absk p_b^+ + e^{2i\Phi} \bm{k} q_b^- - e^{-2i\Phi} \bm{k}^\star r_b^-\right) .
\end{align}
The coefficients are of similar complexity as above, and we refer the reader to \ref{app:coefficientsCross} for details. The distribution of the cross sections $p(\sigma_{a b}\vert\Xi)$ is a Bessel transform of the bivariate characteristic function
\begin{equation}
	p(\sigma_{a b}\vert\Xi) = \frac{1}{4\pi} \int\dd^2\bm{k} J_0\left(\sqrt{\sigma_{a b\myprime}} \lvert \bm{k} \rvert\right) R(\bm{k}\vert\Xi) .
	\end{equation}
With this result we calculate all moments of the cross sections according to
\begin{equation}\label{eqn:momentCrossSections}
	\langle \sigma_{ab}^n \rangle = (-1)^n\sum_{l=0}^{n} \binom{n}{l} \partial_{\text{Re}(\bm{k})}^{2l} \partial_{\text{Im}(\bm{k})}^{2(n-l)} R(\bm{k}) \Bigg\vert_{\bm{k}=0} .
\end{equation}
We list the first moment up to first order in $\Xi$
\begin{align}
	\langle \sigma_{ab} \rangle =& \frac{1}{2} \left(4 p_a^0 p_b^0 + p_a^+ p_b^+ + p_a^- p_b^-\right)\notag\\
	&+ \Xi \Biggl(
		8 \left(4 p_a^0 p_b^0 + p_a^+ p_b^+ + p_a^- p_b^-\right) \left(1 - \cos\theta \cosh\theta_1 \cosh\theta_2 \right) \notag\\
		&+ 8 \left(p_a^+ p_b^0 + p_a^0 p_b^+\right) \sin\theta \sinh\theta_1 \cosh\theta_2  + 8 \left(p_a^- p_b^0 + p_a^0 p_b^-\right) \sin\theta \cosh\theta_1 \sinh\theta_2 \notag\\
		&+ 3 \left(4p_a^0 p_b^0 - p_a^+ p_b^+ - p_a^- p_b^-\right) \sin^2\theta+ 4 \left(2p_a^0 p_b^0 + p_a^+ p_b^+\right) \sinh^2\theta_1 \notag\\
		&+ \left(20 p_a^0 p_b^0 + 3p_a^+ p_b^+ + 7 p_a^- p_b^-\right)\sinh^2\theta_2
	\Biggr) + \mathcal{O}(\Xi^2)
\end{align}
which nicely recover the results from Ref. \cite{NKSG2014} for $\xi=0$. We do not list higher orders, but they are accessible through \cref{eqn:momentCrossSections} which we used to calculate the first moment.

\section{New Avenue for Data Analysis Beyond Detailed Balance}
\label{sec:DataAnalysis}
As is evident from the results of \cref{sec:gsegueScatterMatElem} and \cref{sec:goegueScatterMat}, even when time reversal invariance is broken, the distribution remains invariant under the exchange of the two channels $a\leftrightarrow b$. Hence, on the level of the distribution, detailed balance is always fulfilled. While this prevents us from carrying out an analysis of detailed experiments by comparing the distributions, we gained a much more powerful tool. The exact expressions of the distributions allows us to analyse the strength of time reversal invariance breaking by observing the statistics of a single scattering matrix element $S_{am bm\myprime}$ for the GSE or $S_{ab}$ for the GOE. Thus, it is also possible to carry out data analysis in experiments that were not based on detailed balance. Importantly, our expressions also give access to all moments and consequently improves what had previously been possible, e.g. in Ref. \cite{Dietz2009}, where expressions only existed for the first few moments or correlators. Furthermore, to the best of our knowledge, this is the first access to time reversal invariance breaking in GSE scattering systems.

\section{Conclusions}
\label{sec:conclusion}
Using previous results on the distributions of off-diagonal scattering matrix elements for the three Dyson ensembles \cite{NKSG2014,GG2025}, we succeed in calculating the distributions for ensembles with broken time reversal invariance for GSE and GOE systems. Our explicit expression for the characteristic functions naturally include moments of all orders as well. For GSE systems it was a closed expression only exists when time reversal invariance is weakly broken relative to the mean level spacing. For systems without spin such approximations were not necessary, and we managed to derive a closed expression for arbitrary strengths of time reversal invariance breaking. In both cases we reasonably argued that in the limit of infinitely strong breaking we recover the result for systems with unitary symmetry. 

In the case of symplectic symmetry, we find, as expected, that the breaking of time reversal invariance neither influences the equal distribution of real, imaginary parts and different spin orientations, nor destroys the $a\leftrightarrow b$ and $k \to -k$ symmetries in the integral. In systems with orthogonal symmetry we also find the exchange symmetry $a\leftrightarrow b$ which is already present in both limiting cases. Furthermore, we argued that the existence of this symmetry does not stand in contrast with the experimental observations of Ref. \cite{DFH2009}. Instead, it allows us to investigate time reversal invariance breaking by analysing the statistical properties of a single scattering matrix elements. This constitutes the main result of the present work, as it provides a new access to a data analysis of time reversal invariance breaking without detailed balance experiments. Furthermore, it also opens the door for a much more sophisticated analysis, as the distribution provides access to all moments. Additionally, we find that higher order contributions to the time reversal invariance breaking do not distinguish between real and imaginary part of the distribution. Therefore, restoring the equal distribution in the unitary limit.

With the present derivations we have extended previous studies for GOE systems to the distribution of scattering matrix elements. Using these extended methods, we are hopeful that it is possible to carry out much more detailed analysis of past and future experiments. Furthermore, eliminating the necessity to measure multiple scattering matrix elements our results might pave the way for new experimental realizations. Also, we laid the groundwork to investigate time reversal invariance breaking in \emph{GSE} systems for which to the best of our knowledge did not exist any theoretical framework until now. We considered the case in which Kramers' degeneracy is broken in the unitary limit but for experimental realizations the case of unbroken Kramers' degeneracy is also of large interest. Unfortunately, this case is much more difficult as its supersymmetry formulation is not directly related to that of unbroken time reversal invariance which goes well beyond the scope of this paper, and are objective of ongoing research. Furthermore, studying the reciprocity of the distributions in the case in which it is violated for the scattering matrix elements and investigating if there are cases in which it is also violated for the distribution could lead to interesting discoveries of the underlying mechanisms.

\section*{Acknowledgments}
We thank B. Dietz, S. Köhnes and B. Tomakin for fruitful discussions. This research was funded by the Deutsche Forschungsgemeinschaft (DFG, German Research Foundation) within the project \textit{Stochastic Quantum Scattering -- New Tools, New Aspects}, DFG project number 540160740.
\newpage

\appendix
\section{Parametrization of the Saddle Point Manifold}
\label{app:parametrization}

The saddle point manifold is parametrized by 
\begin{equation}
	Q = - i \mathcal{U}^{-1} \begin{bNiceMatrix}
		\cos\widehat{\theta} & \sin\widehat{\theta} \\
		\sin\widehat{\theta} & - \cos\widehat{\theta}
	\end{bNiceMatrix} \mathcal{U}
\end{equation}
\subsection{Symplectic Case} 
For systems with symplectic symmetry we have
\begin{equation}
	\widehat{\theta} = \diag\left(\widehat{\theta}_{\text{BB}}, \widehat{\theta}_{\text{FF}}\right), \quad \widehat{\theta}_{\text{BB}} = i \theta \mathds{1}_2, \quad \widehat{\theta}_{\text{FF}} = \begin{bNiceMatrix}
		\theta_1 & \theta_2 \\
		\theta_2 & \theta_1
	\end{bNiceMatrix}
\end{equation}
and the singular values fulfill $\theta\geq 0, 0 \leq \theta_1 \leq \pi, 0 \leq \theta_2 \leq \pi/2$. Additionally, the transformations $\mathcal{U} = \diag\left(\mathcal{U}_1, \mathcal{U}_2\right)$ with $\mathcal{U}_1=\widehat{U}^\dagger u_1, \mathcal{U}_2=u_2$ are
\begin{align}
	\widehat{U} =& U \oplus \mathds{1}_2, \quad U = \begin{bNiceMatrix}
		u e^{i\varphi_1} & \sqrt{1-u^2} e^{i\varphi_2} \\
		-\sqrt{1-u^2} e^{-i\varphi_2} & u e^{-i\varphi_1}
	\end{bNiceMatrix} \notag\\
	u_j =& O_j v_j, \quad O_j = \mathds{1}_2 \oplus \diag\left(e^{-i\phi_j}, e^{+i\phi_j}\right) \notag\\
	v_j =& \exp\left(i^{j-1} \left(Y - (-1)^j \frac{Y^3}{3}\right)\right), \quad Y_j = \begin{bNiceMatrix}
		0 & - \xi_j^\dagger \\
		\xi_j & 0
	\end{bNiceMatrix}, \notag\\
	\xi_j =& \begin{bNiceMatrix}
		\mu_j & \nu_j^\star \\
		\nu_j & \mu_j^\star
	\end{bNiceMatrix}
\end{align}
where $u\in[0,1], \varphi_j\in[0,2\pi], \phi_j\in[0,2\pi]$, $\mu_j, \nu_j$ and their complex conjugates are anticommuting variables and $j=1,2$. The corresponding volume element is
\begin{align}
	\dd\mu(Q) =& \mathcal{B} \dd\theta \dd\theta_1 \dd\theta_2 \dd u \dd\varphi_1 \dd\varphi_2 \dd\phi_1 \dd\phi_2 \dd[\Upsilon], \notag\\
	\mathcal{B} =& \frac{2 u \sin\theta_1 \sin\theta_2 \sinh^3\theta}{\left(\cos\left(\theta_1 + \theta_2\right) - \cosh\theta\right)^2 \left(\cos\left(\theta_1 - \theta_2\right) - \cosh\theta\right)^2}
\end{align}
where $\dd[\Upsilon]$ contains all differentials of anticommuting variables.

\subsection{Orthogonal Case}
For systems with orthogonal symmetry we have
\begin{equation}
	\widehat{\theta} = \diag\left(\widehat{\theta}_{\text{BB}}, \widehat{\theta}_{\text{FF}}\right), \quad \widehat{\theta}_{\text{BB}} = \begin{bNiceMatrix}
		\theta_1 & \theta_2 \\
		\theta_2 & \theta_1
	\end{bNiceMatrix}, \quad \widehat{\theta}_{\text{FF}} = \theta \mathds{1}_2
\end{equation}
and the singular values fulfill $0\leq\theta\leq \pi, \theta_1,\theta_2 \geq 0$. Additionally, the transformations $\mathcal{U} = \diag\left(\mathcal{U}_1, \mathcal{U}_2\right)$ with $\mathcal{U}_1=\widehat{U}^\dagger u_1, \mathcal{U}_2=u_2$ are
\begin{align}
	\widehat{U} =&  \mathds{1}_2 \oplus U, \quad U = \begin{bNiceMatrix}
		u e^{i\varphi_1} & \sqrt{1-u^2} e^{i\varphi_2} \\
		-\sqrt{1-u^2} e^{-i\varphi_2} & u e^{-i\varphi_1}
	\end{bNiceMatrix} \notag\\
	u_j =& O_j v_j, \quad O_j = \diag\left(e^{-i\phi_j}, e^{+i\phi_j}\right) \oplus \mathds{1}_2 \notag\\
	v_j =& \exp\left(i^{j-1} \left(Y - (-1)^j \frac{Y^3}{3}\right)\right), \quad Y_j = \begin{bNiceMatrix}
		0 & - \xi_j^\dagger \\
		\xi_j & 0
	\end{bNiceMatrix}, \notag\\
	\xi_j =& \begin{bNiceMatrix}
		\mu_j & \nu_j \\
		\nu_j^\star & \mu_j^\star
	\end{bNiceMatrix}
\end{align}
where $u\in[0,1], \varphi_j\in[0,2\pi], \phi_j\in[0,2\pi]$, $\mu_j, \nu_j$ and their complex conjugates are anticommuting variables and $j=1,2$. The corresponding volume element is
\begin{align}
	\dd\mu(Q) =& \mathcal{B} \dd\theta \dd\theta_1 \dd\theta_2 \dd u \dd\varphi_1 \dd\varphi_2 \dd\phi_1 \dd\phi_2 \dd[\Upsilon], \notag\\
	\mathcal{B} =& \frac{2 u \sinh\theta_1 \sinh\theta_2 \sin^3\theta}{\left(\cosh\left(\theta_1 + \theta_2\right) - \cos\theta\right)^2 \left(\cosh\left(\theta_1 - \theta_2\right) - \cos\theta\right)^2}
\end{align}
where $\dd[\Upsilon]$ contains all differentials of anticommuting variables.

\section{Coefficients for Time Reversal Invariance Breaking in GSE}
\label{app:coefficientsGSE}
We introduce the notation
\begin{equation}
	\left\{f(a,b)\right\} = f(a,b) + f(b,a).
\end{equation}
\newpage
The coefficients in \cref{eqn:cfIB} are
\begin{align}
	\bm{\iota}_0 =& \frac{1}{\omega_{ab}^2} \left\{\iota_{0,ab}\right\}, \notag\\
	\iota_{0,c c\myprime} =& -4 \omega_{ab}^2 \sin\left(\theta_1 + \theta_2\right)\sin\left(\theta_1 - \theta_2\right)\notag\\
	&- \left(8 G_{c c\myprime}^2 t_{cc } + 4t_c^+ t_{c\myprime}^- - \frac{1}{2}k^2 t_{cc} t_{c\myprime c\myprime}\right)\sinh^2\theta \notag\\
	&+ 2 i \kappa_{0c} \sinh\theta \notag\\
	&\quad\times \left(\left(3t_{c\myprime}^- - t_{c\myprime}^+ \right) \sin\left(\theta_1 + \theta_2\right) + \left(3 t_{c\myprime}^+ - t_{c\myprime}^-\right) \sin\left(\theta_1 - \theta_2\right)\right)\notag\\
	&+ \upsilon_1 \omega_{c c\myprime} \left(12 - k^2 \left(t_{cc} + \omega_{c c\myprime}^2\right)\right)G_{c c\myprime} \left(t_c^+ + t_c^-\right)\notag\\
	&+\omega_{c c\myprime}^2 k^2 \left(\frac{1}{2}t_{c c\myprime} + \omega_{c c\myprime}^2\right) \notag\\
	&\quad\times\left( \cosh\theta \left(1-\cos\left(\theta_1 + \theta_2\right) -\cos\left(\theta_1 - \theta_2\right)\right) + \cos^2\theta_2\right)\notag\\
	&-\frac{1}{2}\omega_{c c\myprime}^4 k^2 \sinh^2\theta 
\end{align}
and
\begin{align}
	\bm{\iota}_1 =& \frac{1}{2\omega_{ab}^2} \left\{\iota_{1,ab}\right\}, \notag\\
	\iota_{1,c c\myprime} =& -2 \sinh^2\theta\Biggl[\left(2G_{c c\myprime}^2 t_{cc} + t_c^+ t_{c\myprime}^-\right)\left(-8 + \omega_{c c\myprime}^2k^2\right)  \notag\\*
	&\bralign{-2\sinh^2\theta}+ k^2 \left( t_{c c} t_{c\myprime c\myprime} + \frac{1}{2}\omega_{c c\myprime}^2\left(2\omega_{c c\myprime}^2 - t_{c c\myprime}\right)\right)\Biggr] \notag\\*
	&- 4 i \left(4 - \omega_{c c\myprime}^2 k^2\right) \kappa_{0c} \sinh\theta \notag\\*
	&\quad\times\left(\left(3t_{c\myprime}^- - t_{c\myprime}^+ \right) \sin\left(\theta_1 + \theta_2\right) + \left(3 t_{c\myprime}^+ - t_{c\myprime}^-\right) \sin\left(\theta_1 - \theta_2\right)\right) \notag\\*
	&- 2 \upsilon_1 \omega_{c c\myprime} G_{c c\myprime} \left(12 - k^2\left(t_{c c} + 2\omega_{c c\myprime}^2\right)\right) \left(t_c^+ + t_c^-\right) \notag\\*
	&\quad\times \left(\sin\left(\theta_1 + \theta_2\right) + \sin\left(\theta_1 - \theta_2\right)\right) \notag\\*
	&+ 2 \omega_{c c\myprime}^2 \left(16 - k^2\left(4\omega_{c c\myprime}^2 + \left(t_c^+ + t_c^-\right)\left(t_{c\myprime}^+ + t_{c\myprime}^-\right)\right)\right) \notag\\*
	&\quad\times \sin\left(\theta_1 + \theta_2\right) \sin\left(\theta_1 - \theta_2\right) \notag\\*
	&- \omega_{c c\myprime}^2k^4 \left(t_{c c} t_{c\myprime c\myprime} + \omega_{c c\myprime}^2 \left(t_{c c\myprime} + \omega_{c c\myprime}^2\right)\right) \sin^2\theta_2.
\end{align}

\section{Coefficients of the Bessel Functions for the Distribution of Scattering Matrix Elements in GOE Systems}
\label{app:coefficients}
When exchanging $a \leftrightarrow b$ and setting $\Phi \to -\Phi+\pi/2$ we have $\mathcal{X} \to (-1)^s\mathcal{Y}$ and vice versa. We use this property to significantly reduce the length of our expressions. Unfortunately, they turn out to still be rather long. We introduce the bracket
\begin{equation}
	\{f(a,b,\Phi)\}_+ = f(a,b,\Phi) + f(b,a,-\Phi+\pi/2) .
\end{equation}
The coefficients are
\begin{flalign}
	\bm{\iota}_1 =& \frac{k^2}{2048} \left(\iota_{00} + \left\{\iota_{02} e^{2i\Phi} + \iota_{04} e^{4i\Phi}  + \iota_{06} e^{6i\Phi} + \iota_{08} e^{8i\Phi}\right\}_+ \right),&&\\
	\iota_{00} =&  \mathfrak{u}_1 
	\Biggl[
	k^2 
	\Biggl(
	\left(32 \left(p_a^0\right)^2 + 8\left(p_a^-\right)^2 \right) \left(4 \left(p_b^0\right)^2 + \left(p_b^-\right)^2 + q_b^+ r_b^+\right)  &&\notag\\*
	&\bralign{\mathfrak{u}_1}\bralign{k^2}+ 8q_a^+ r_a^+\left( 4 \left(p_b^0\right)^2  + \left(p_b^-\right)^2 \right) + 2q_b^- r_b^- \left( \left(p_a^+\right)^2 + q_a^- r_a^-\right)  &&\notag\\*
	&\bralign{\mathfrak{u}_1}\bralign{k^2}+ \left(p_b^+\right)^2\left(3 \left(p_a^+\right)^2 + 2  q_a^- r_a^-\right) &&\notag\\*
	&\bralign{\mathfrak{u}_1}\bralign{k^2}- 4 p_a^- p_a^+ \left(2p_b^- p_b^+ +  q_b^+ r_b^- + q_b^- r_b^+\right)  &&\notag\\*
	&\bralign{\mathfrak{u}_1}\bralign{k^2}- 4 p_b^- p_b^+ \left(q_a^+ r_a^- - q_a^- r_a^+ \right) - 4 r_a^+ r_b^+ \left( q_a^- q_b^- - 2 q_a^+ q_b^+ \right)&&\notag\\*
	&\bralign{\mathfrak{u}_1}\bralign{k^2}- 4 q_a^+ q_b^+ r_a^- r_b^- &&\notag\\*
	&\bralign{\mathfrak{u}_1}\bralign{k^2}+ 2 (-1)^s \left(-2 p_a^- p_b^- + p_a^+ p_b^+\right) &&\notag\\*
	&\bralign{\mathfrak{u}_1}\bralign{k^2}\quad \times\left(q_a^- q_b^-  - 2 q_a^+ q_b^+ + r_a^- r_b^- - 2 	r_a^+ r_b^+\right)
	\Biggr) &&\notag\\*
	&\bralign{\mathfrak{u}_1}+ 256 p_a^- \left(-2\Xi \chi_1  p_b^0 + p_b^- \left(-1 + \Xi \upsilon_1\right)\right) &&\notag\\*
	&\bralign{\mathfrak{u}_1}+ 32 p_a^0 \left(-16 \Xi \chi_1 p_b^-  + p_b^0 \left(96 + k^2 p_a^+ p_b^+ - 32 \Xi \upsilon_1\right)\right)&&\notag\\*
	&\bralign{\mathfrak{u}_1}+ 16 p_a^+ p_b^+ \left(-7 + 4 \Xi \left(-\upsilon_2 - \Xi \left(\upsilon_1^2 + \chi_1^2\right)\right)\right)
	\Biggr] &&\notag\\*
	&-64 \mathfrak{u}_3
	\Biggl[
	p_a^0 \left(-8\Xi \chi_1 p_b^-  + 7\Xi \chi_2 p_b^+ \right) &&\notag\\*
	&\bralign{-64 \mathcal{u}_3}+ 8 p_a^0 p_b^0 \left(16 - 4\Xi \upsilon_7 + \Xi^2 \left(\upsilon_1^2 + \chi_1^2\right)\right) &&\notag\\*
	&\bralign{-64 \mathcal{u}_3} +\Xi 
	\Biggl(
	16 p_b^0 \sin\theta &&\notag\\*
	&\bralign{-64 \mathcal{u}_3}\bralign{+\Xi}\quad\times\left(-8 p_a^- \cosh\theta_1 \sinh\theta_2+ 7 p_a^+ \sinh\theta_1 \cosh\theta_2\right)  &&\notag\\*
	&\bralign{-64 \mathcal{u}_3}\bralign{+\Xi} +  p_a^+ p_b^+ \left(48 \sin^2\theta - \Xi\left(32 \sin^2\theta \upsilon_1 - \chi_2^2 + \chi_1^2\right)\right) &&\notag\\*
	&\bralign{-64 \mathcal{u}_3}\bralign{+\Xi} + 64 \sin^2\theta p_a^- p_b^- 
	\Biggr)		
	\Biggr] &&\notag\\*
	&+ 512 \Xi \mathfrak{u}_5 
	\Biggl[
	p_a^0 p_b^0 \left(160 \sin^2\theta -\Xi \left(32 \sin^2\theta \upsilon_1 - \chi_2^2 + \chi_1^2\right)\right) &&\notag\\*
	&\bralign{512\Xi\mathfrak{u}_5}+ 32\Xi p_a^+ p_b^+ \sin^2\theta
	\Biggr] &&\notag\\*
	&+ 131072 \Xi^2 \mathfrak{u}_7 p_a^0 p_b^0 \sin^4\theta &&\\
	\iota_{02} =& i \left(\mathfrak{j}_{02,ab} + \mathfrak{j}_{02,ba}^\star\right),&&\\
	\iota_{04} =& \frac{1}{2} \mathfrak{u}_1 
	\Biggl[
	-k^2 
	\Biggl(
	(-1)^s \left(\left(p_a^+ q_b^-\right)^2 + \left(p_b^+ r_a^-\right)^2\right) &&\notag\\*
	&\bralign{\frac{1}{2} \mathfrak{u}_1}\bralign{-k^2}+ 8 
	\Biggl(
	q_b^- r_a^- \left(4 p_a^0 p_b^0  + p_a^+ p_b^+ \right) - p_a^- p_b^+ q_b^+ r_a^- &&\notag\\*
	&\bralign{\frac{1}{2} \mathfrak{u}_1}\bralign{-k^2}\bralign{+8}- p_a^+ p_b^- q_b^- r_a^+ + 2 p_a^- p_b^- q_b^+ r_a^+
	\Biggr)
	\Biggr) &&\notag\\*
	&\bralign{\frac{1}{2}\mathfrak{u}_1}+ 64 q_b^- r_a^- \left(1- 2 \Xi \upsilon_3 + \Xi^2\left(\upsilon_1^2 + \chi_1^2\right)\right)
	\Biggr] &&\notag\\*
	&- 32\Xi \mathfrak{u}_3 q_b^- r_a^- \left(-64 \sin^2\theta + \Xi\left(32 \sin^2\theta \upsilon_1 - \chi_2^2 + \chi_1^2 \right)\right) &&\notag\\*
	&+ 8192 \Xi^2 \mathfrak{u}_5 q_b^- r_a^- \sin^4\theta ,&&\\
	\iota_{06} =& -i \mathfrak{u}_1 q_b^- r_a^- \left(i^s p_a^+ q_b^- + (-i)^s p_b^+ r_a^-\right),&&\\
	\iota_{08} =& \frac{1}{2} \mathfrak{u}_1 \left(q_b^- r_a^-\right)^2,&&\\
	\mathfrak{j}_{02,c c\myprime} =& i^s\mathfrak{u}_1 \Biggl[
	k^2 \Biggl( 4p_c^+  q_{c\myprime}^- \left(4 p_c^0p_{c\myprime}^0 - p_c^- p_{c\myprime}^-\right) + 8p_{c\myprime}^- q_{c\myprime}^+\left(4 \left(p_c^0\right)^2  + \left(p_c^-\right)^2 \right)&&\notag\\
	&\bralign{i^s \mathfrak{u}_1}+ p_c^+ q_{c\myprime}^+ \left(3p_c^+ q_{c\myprime}^- - 4 p_c^- p_{c\myprime}^+\right) &&\notag\\
	&\bralign{i^s \mathfrak{u}_1}+ 2\left(r_c^- p_{c\myprime}^+ -2 r_c^+ p_{c\myprime}^-\right) \left(-2 q_c^+ q_{c\myprime}^+ + q_c^- q_{c\myprime}^- \right)
	\Biggr) &&\notag\\
	&\bralign{i^s \mathfrak{u}_1}+32 p_c^- q_{c\myprime}^+ \left(-3 + 2 \Xi \upsilon_1\right)&&\notag\\
	&\bralign{i^s \mathfrak{u}_1}+ 16 \Xi \left(-8 \chi_1 p_c^0 q_{c\myprime}^+ - p_c^+ q_{c\myprime}^- \left(-\upsilon_4 + 2 \Xi\left(\upsilon_1^2 + \chi_1^2\right)\right)\right)
	\Biggr]&&\notag\\
	&+ 32 i^s \Xi \mathfrak{u}_3 \Biggl[
	32p_c^0 \sin\theta \left(- q_{c\myprime}^- \sinh\theta_1 \cosh\theta_2  + 2  q_{c\myprime}^+ \cosh\theta_1 \sinh\theta_2\right) &&\notag\\
	&\bralign{+32 i^s \Xi \mathfrak{u}_3} + p_c^+ q_{c\myprime}^- \left(-56\sin\theta + \Xi \left(32 \sin\theta \upsilon_1 -\chi_2^2 + \chi_1^2\right)\right) &&\notag\\
	&\bralign{+32 i^s \Xi \mathfrak{u}_3}- 32 \sin\theta p_c^- q_{c\myprime}^+
	\Biggr]&&\notag\\
	&+ 8192 i^s \Xi^2 \mathfrak{u}_5 p_c^+ q_{c\myprime}^- \sin^4\theta,&&
\end{flalign}

\begin{flalign}
	\bm{\iota}_1 =&  \frac{-ik}{1024 \sqrt{\mathcal{X} \mathcal{Y}}}\left\{\iota_{10} + \iota_{12} e^{2i\Phi} + \iota_{14} e^{4i\Phi} + \iota_{16} e^{6i\Phi} \right\}_+ ,&&\\
	\iota_{10} =& 2 i^{s} \mathfrak{j}_{10,ba} \mathcal{X} ,&&\notag\\
	\iota_{12} =& 2i(-1)^s k^2  \left(\mathfrak{j}_{12,ba} \mathcal{X} + \mathfrak{j}_{12,ab} \mathcal{Y}\right),&&\notag\\
	\iota_{14} =& i^s k^2 \left(\mathfrak{j}_{14,ba} \mathcal{X} + (-1)^s\mathfrak{j}_{14,ab}^\star \mathcal{Y}\right),&&\notag\\
	\iota_{16} =& 2i(-1)^{s+1} k^2  \left(\left(-1+\Xi \upsilon_1\right) \mathfrak{u}_1 - 4 \kappa_5 \Xi \mathfrak{u}_3\right)\left(\mathfrak{j}_{16,ba} \mathcal{X} + \mathfrak{j}_{16,ab}^\star \mathcal{Y}\right),&&\\
	\mathfrak{j}_{10,c c\myprime} =&\mathfrak{u}_1 \Biggl[
	-8 p_b^+ \left(9 + 4 \Xi \left(- \upsilon_5 + \Xi \left(\upsilon_1^2 + \chi_1^2\right)\right)\right) &&\notag\\*
	&\bralign{\mathfrak{u}_1}+ k^2 
	\Biggl(
	(-1)^s p_b^+ \left(2 q_b^+ q_b^+ - q_a^- q_b^- + 2 r_a^+ r_b^+ - r_a^- r_b^- \right) \left(3 - 2 \Xi \upsilon_1\right) &&\notag\\*
	&\bralign{\mathfrak{u}_1}\bralign{+k^2}+ p_a^+ \left(32 \left(p_b^0\right)^2 + 8 \left(p_b^-\right)^2 - 3 \left(p_b^+\right)^2 + 8 q_b^+ r_b^+ - 2 q_b^- r_b^-\right) &&\notag\\*
	&\bralign{\mathfrak{u}_1}\bralign{+k^2}+ 2 \Xi \upsilon_1 p_a^+\left(3 \left(p_b^+\right)^2 + 2 q_b^- r_b^-\right) &&\notag\\*
	&\bralign{\mathfrak{u}_1}\bralign{+k^2}- 4 p_a^-\left( p_b^+ \left(p_b^- \left(1+ 2\Xi \upsilon_1\right) -4\Xi \chi_1 p_b^0 \right)\right) &&\notag\\*
	&\bralign{\mathfrak{u}_1}\bralign{+k^2}-4 \Xi\upsilon_1 p_a^- \left(q_b^+ r_b^- + q_b^- r_b^+\right) &&\notag\\*
	&\bralign{\mathfrak{u}_1}\bralign{+k^2}+ \Xi p_a^0  \left(\chi_1 \left(2 p_b^- p_b^+ + q_b^+ r_b^- + q_b^- r_b^+\right)\right) &&\notag\\*
	&\bralign{\mathfrak{u}_1}\bralign{+k^2}+ 2\Xi p_a^0 p_b^0 p_b^+ \left(-3 + 2 \Xi \upsilon_1\right)
	\Biggr)
	\Biggr]
	&&\notag\\*
	&+ 32 (-1)^s \Xi k^2 \mathfrak{u}_3 p_b^+ \left(2 q_a^+ q_b^+ -q_a^- q_b^- + 2 r_a^+ r_b^+ - r_a^- r_b^- \right)\sin\theta  
	&&\notag\\*
	&+4 \Xi \mathfrak{u}_3 
	\Biggl[
	4\sin^2\theta \Biggl(
	p_b^+ \left(-160 - 32k^2 p_a^0 p_b^0\right) &&\notag\\*
	&\bralign{+4\Xi\mathfrak{u}_3}\bralign{4\sin^2\theta}- 2k^2 p_a^+ \left(3 \left(p_b^+\right)^2 + 2 q_b^- r_b^-\right) &&\notag\\*
	&\bralign{+4\Xi\mathfrak{u}_3}\bralign{4\sin^2\theta}+ 4k^2 p_a^- \left(2 p_b^+ p_b^- + q_b^+ r_b^- + q_b^- r_b^+\right)
	\Biggr) &&\notag\\*
	&\bralign{+4\Xi\mathcal{u}_3}+ \chi_1 \Biggl(
	-4k^2 p_a^- p_b^0 p_b^+ -2k^2 p_a^0 \left(2 p_b^+ p_b^- + q_b^+ r_b^- + q_b^- r_b^+\right)
	\Biggr) &&\notag\\*
	&\bralign{+4\Xi\mathcal{u}_3}+ \chi_2 \Biggl(
	k^2 p_a^0 \Biggl(16 \left(p_b^0\right)^2 + \left(p_b^+\right)^2 + 4\left(p_b^-\right)^2 &&\notag\\*
	&\bralign{+4\Xi\mathcal{u}_3}\bralign{+\chi_2}\bralign{k^2 p_a^0}+ 4 q_b^+ r_b^+ + q_b^- r_b^-\Biggr) &&\notag\\*
	&\bralign{+4\Xi\mathcal{u}_3}\bralign{+\chi_2}+2 p_b^0 \left(-8 + k^2 p_a^+ p_b^+\right)
	\Biggr) &&\notag\\*
	&\bralign{+4\Xi\mathcal{u}_3}+ 8 p_b^+ \Xi \left(8 \kappa_5 \upsilon_1 + \chi_1^2 - \chi_2^2\right)
	\Biggr] &&\notag\\*
	&- 8192 \Xi^2 \mathfrak{u}_5 p_b^+ \sin^4\theta,&&\\
	\mathfrak{j}_{12,c c\myprime} =& \mathfrak{u}_1 
	\Biggl[
	q_c^- 
	\Biggl(
	2p_c^+ p_{c\myprime}^+ \left(-2 + 3 \Xi \upsilon_1\right) -4 p_c^- p_{c\myprime}^- \left(1 + \Xi \upsilon_1\right) &&\notag\\*
	&\bralign{\mathfrak{u}_1}\bralign{q_c^-}+ 8\Xi\chi_1 \left(p_c^0 p_{c\myprime}^- + p_c^- p_{c\myprime}^0 + 2 p_c^0 p_{c\myprime}^0 \left(-2 + \Xi \upsilon_1\right)\right)
	\Biggr) &&\notag\\*
	&\bralign{\mathfrak{u}_1}+ 2q_c^+ p_c^+ p_{c\myprime}^- \left(1-2 \Xi \upsilon_1\right) + 8 p_{c\myprime}^+ p_c^- q_c^+ &&\notag\\*
	&\bralign{\mathfrak{u}_1}+ 8\Xi \chi_1 p_{c\myprime}^0  p_c^+ q_c^+ 
	&&\notag\\*
	&\bralign{\mathfrak{u}_1}+ (-1)^s 
	\Biggl(
	2 q_c^- \left( r_c^- r_{c\myprime}^- - 2 r_c^+ r_{c\myprime}^+ \right) \left(-1 + \upsilon_1 \Xi\right) &&\notag\\*
	&\bralign{\mathfrak{u}_1}\bralign{+(-1)^s}+ \left(p_c^+\right)^2 r_{c\myprime}^- \left(-2 + 3 \upsilon_1 \Xi\right) &&\notag\\*
	&\bralign{\mathfrak{u}_1}\bralign{+(-1)^s}+ 2 p_c^+ r_{c\myprime}^+ \left(p_c^- \left(1-2\Xi \upsilon_1\right) + 4\Xi p_c^0 \chi_1\right)
	\Biggr)
	\Biggr] &&\notag\\*
	&+ 64 \Xi \mathfrak{u}_3 \sin\theta \Biggr[
	q_c^- p_c^0 \left(p_{c\myprime}^+ \sinh\theta_1 \cosh\theta_2 - 2 p_{c\myprime}^- \cosh\theta_1 \sinh\theta_2\right) &&\notag\\*
	&\bralign{+16\Xi \mathfrak{u}_3 \sin\theta}+ q_c^-p_{c\myprime}^0 \left(p_{c}^+ \sinh\theta_1 \cosh\theta_2 - 2 p_{c}^- \cosh\theta_1 \sinh\theta_2\right) &&\notag\\*
	&\bralign{+16\Xi \mathfrak{u}_3 \sin\theta}+ 2 q_c^+ p_{c\myprime}^0 \left(2p_c^- \sinh\theta_1 \cosh\theta_2 - p_c^+ \cosh\theta_1 \sinh\theta_2\right) &&\notag\\*
	&\bralign{+16\Xi \mathfrak{u}_3\sin\theta }+ (-1)^s p_c^0 p_c^+ \left(r_{c\myprime}^-\sinh\theta_1 \cosh\theta_2 - 2 r_{c\myprime}^+ \cosh\theta_1 \sinh\theta_2\right) &&\notag\\*
	&\bralign{+16\Xi \mathfrak{u}_3\sin\theta}+ \frac{1}{4}\sin\theta  
	\Biggl(
	2q_c^-\left(-8 p_c^0 p_{c\myprime}^0  + 2 p_c^- p_{c\myprime}^- - 3 p_c^+ p_{c\myprime}^+ \right) &&\notag\\*
	&\bralign{+16\Xi \mathfrak{u}_3\sin\theta}\bralign{+4\sin^2\theta}+ 4 p_c^+ q_c^+ p_{c\myprime}^-  &&\notag\\*
	&\bralign{+16\Xi \mathfrak{u}_3\sin\theta}\bralign{+4\sin^2\theta}+ (-1)^s p_c^+\left(-3 p_c^+ r_{c\myprime}^- + 4 p_c^- r_{c\myprime}^+\right) &&\notag\\*
	&\bralign{+16\Xi \mathfrak{u}_3\sin\theta}\bralign{+4\sin^2\theta}- 2 (-1)^s q_c^- \left( r_c^- r_{c\myprime}^-  - 4 r_c^+ r_{c\myprime}^+ \right)
	\Biggr)
	\Biggl],&&\\
	\mathfrak{j}_{14,c c\myprime} =& \mathfrak{u}_1 q_c^- 
	\Biggl[
	(-1)^s p_{c\myprime}^+ q_c^- \left(3 - 2\Xi \upsilon_1\right) - 4 p_c^+ r_{c\myprime}^- \left(-1 + 2 \Xi \upsilon_1\right)&&\notag\\
	&\bralign{\mathfrak{u}_1 q_c^-}- 8\Xi \left(2 p_c^0 \chi_1 - \upsilon_1 p_c^- \right) r_{c\myprime}^+
	\Biggr] &&\notag\\
	&+ 32 \Xi \mathfrak{u}_3 q_c^- \sin\theta
	\Biggl[
	(-1)^s\sin\theta p_{c\myprime}^+ q_c^- &&\notag\\
	&\bralign{+ 32 \Xi \mathfrak{u}_3 q_c^- \sin\theta}+ 4 \left(\sin\theta p_c^+ - \sinh\theta_1 \cosh\theta_2 p_c^0\right) r_{c\myprime}^- &&\notag\\
	&\bralign{+ 32 \Xi \mathfrak{u}_3 q_c^- \sin\theta}- \left(\sin\theta q_c^- - 2 \cosh\theta_1 \sinh\theta_2 p_c^0\right) r_{c\myprime}^+
	\Biggr] ,&&\\
	\mathfrak{j}_{16,c c\myprime} =& r_{c\myprime}^- \left(q_c^-\right)^2,&&
\end{flalign}

\begin{flalign}
	\bm{\iota}_2 =& \frac{k^2}{2048 \mathcal{X} \mathcal{Y}} \left\{\iota_{20} + \iota_{22} e^{2i\Phi} + \iota_{24} e^{4i\Phi} + \iota_{26} e^{6i\Phi}\right\}_+ ,&&\notag\\
	\iota_{20} =& (-1)^s \mathfrak{j}_{20,ba} \mathcal{X}^2 ,&&\notag\\
	\iota_{22} =& i (-i)^{s}\left( \mathfrak{j}_{22,ba} \mathcal{X}^2 + (-1)^s \mathfrak{j}_{22,ab}^\star \mathcal{Y}^2 \right),&&\notag\\
	\iota_{24} =& -k^2 \mathfrak{u}_1 \left( \mathfrak{j}_{24,ba} \mathcal{X}^2 +\mathfrak{j}_{24,ab}^\star \mathcal{Y}^2 \right),&&\notag\\
	\iota_{26} =& (-i)^{s+1} \mathfrak{u}_1\left( \mathfrak{j}_{26,ba} \mathcal{X}^2 + (-1)^s\mathfrak{j}_{26,ab}^\star \mathcal{Y}^2 \right),&&\\
	\mathfrak{j}_{20,c c\myprime} =& \mathfrak{u}_1 
	\Biggl[
	2 k^2 p_b^+ \left(p_a^+ \left(\left(p_b^+\right)^2 + 2 q_b^- r_b^-\right) -2 p_a^- \left(q_b^+ r_b^- + q_b^- r_b^+\right)\right) &&\notag\\
	&\bralign{\mathfrak{u}_1}+ 16 \left(\left(p_b^+\right)^2 + q_b^- r_b^-\right) \left(k^2 p_a^0 p_b^0 - 2 \Xi^2 \left(\upsilon_1^2 + \chi_1^2\right)\right)&&\notag\\
	&\bralign{\mathfrak{u}_1}+ 32 q_b^- r_b^- \left(-1 + 2 \Xi \upsilon_3\right) &&\notag\\
	&\bralign{\mathfrak{u}_1}+ \left(p_b^+\right)^2 \left(8 - 4 p_a^- p_b^- - 32 \Xi \upsilon_2 \right) &&\notag\\
	&\bralign{\mathfrak{u}_1}- (-1)^s \left(p_b^+\right)^2 \left(2 q_a^+ q_b^+ - q_a^- q_b^- + 2 r_a^+ r_b^+ - r_a^- r_b^- \right)
	\Biggr] &&\notag\\
	&+ 32 \Xi \mathfrak{u}_3
	\Biggl(
	-16 \sin\theta\left(6 \sinh\theta_1 \cosh\theta_2 p_b^0 p_b^+ + \sin\theta q_b^- r_b^-\right) &&\notag\\
	&\bralign{32\Xi\mathfrak{u}_3}+ \left( \left(p_b^+\right)^2 + q_b^- r_b^-\right) &&\notag\\
	&\bralign{32\Xi\mathfrak{u}_3}\quad\times\left(-48 \sin^2\theta+\Xi \left(32 \sin^2\theta \upsilon_1 - \chi_2^2 + \chi_1^2\right)\right) 
	\Biggl) &&\notag\\
	&- 8192\Xi^2 \mathfrak{u}_5 \sin^4\theta  \left(\left(p_b^+\right)^2 + q_b^- r_b^-\right) ,&&\\
	\mathfrak{j}_{22,c c\myprime} =& \mathfrak{u}_1 
	k^2 
	\Biggl[
	(-1)^s \left(\left(p_c^+\right)^2 \left(p_c^+ r_{c\myprime}^- - 2 p_c^- r_{c\myprime}^+\right) + 2 p_c^+ q_c^- \left(r_{c\myprime}^- r_c^- - 2 r_{c\myprime}^+ r_c^+\right)\right) &&\notag\\
	&\bralign{\mathfrak{u}_1 k^2}+ 16 p_{c\myprime}^0 p_c^0 p_c^+ q_c^- - p_{c\myprime}^+ q_c^- \left(3 \left(p_c^+\right)^2 + q_c^- r_c^-\right) &&\notag\\
	&\bralign{\mathfrak{u}_1 k^2 }- 2 p_{c\myprime}^- \left(2 p_c^- p_c^+ q_c^- + \left(p_c^+\right)^2 q_c^+ + \left(q_c^-\right)^2 r_c^+\right)
	\Biggr] &&\notag\\
	&- 16 \mathfrak{u}_1 p_c^+ q_c^- \Xi \left(-\upsilon_4 + 2 \Xi \left(\upsilon_1^2 + \chi_1^2\right)\right) &&\notag\\
	&- 256 \Xi  \mathfrak{u}_3 q_c^-\sin\theta \left(2 \sinh\theta_1 \cosh\theta_2 p_c^0 + \sin\theta p_c^+ \left(7 - \Xi \upsilon_8 \right)\right)&&\notag\\
	&- 8192 \Xi^2 \mathfrak{u}_5 q_c^- p_c^+ \sin^4\theta,&&\\
	\mathfrak{j}_{24,cc\myprime} =&  \left( p_c^+ p_{c\myprime}^+ - 2 p_c^- p_{c\myprime}^-\right) \left(q_c^-\right)^2 &&\notag\\
	& +(-1)^s q_c^- \left(2 p_c^+ \left( p_c^+ r_{c\myprime}^- - 2 p_c^- r_{c\myprime}^+\right) + q_c^-\left( r_c^- r_{c\myprime}^- - 2 r_c^+ r_{c\myprime}^+ \right)\right),&&\\
	\mathfrak{j}_{26,cc\myprime} =& \left(p_c^+ r_{c\myprime}^- - 2 p_c^- r_{c\myprime}^+\right) \left(q_c^-\right)^2,&&
\end{flalign}

\begin{flalign}
	\bm{\iota}_3 =& i\frac{k^3 \sqrt{\mathcal{X} \mathcal{Y}}}{1024 \mathcal{X}^2 \mathcal{Y}^2} \left\{\iota_{30}  + \iota_{32} e^{2i\Phi} + \iota_{34} e^{4i\Phi}\right\}_+,&&\notag\\
	\iota_{30} =& 2 (-i)^s \mathcal{X}^3 \mathfrak{j}_{30,b} ,&&\notag\\
	\iota_{32} =& 2i \left(\mathfrak{j}_{32,b} \mathcal{X}^3 + \mathfrak{j}_{32,a}^\star \mathcal{Y}^3\right), &&\notag\\
	\iota_{34} =& -i^s \left(\left(-3 + 2 \Xi \upsilon_1\right) \mathfrak{u}_1 - 32 \Xi \mathfrak{u}_3 \sin^2\theta\right) \left(\mathfrak{j}_{34,b} \mathcal{X}^3 + (-1)^s \mathfrak{j}_{34,a}^\star \mathcal{Y}^3\right) ,&&\\
	\mathfrak{j}_{30,c} =& \mathfrak{u}_1 \left(-1+2 \Xi \upsilon_1\right) p_c^+ \left(\left(p_c^+\right)^2 + 2 q_c^- r_c^-\right) &&\notag\\*
	&+ 16 \Xi \mathfrak{u}_3 \sin\theta \Bigl(\left(4 \sinh\theta_1 \cosh\theta_2 p_c^0 - 2 \sin\theta p_c^+ \right) \left(\left(p_c^+\right)^2 + q_c^- r_c^-\right) &&\notag\\*
	&\bralign{+16\Xi\mathcal{u}_3 \sin\theta}- 2 \sin\theta p_c^+  q_c^- r_c^-\Bigr) ,&&\\
	\mathfrak{j}_{32,c} =& \mathfrak{u}_1q_c^- \left((-1 + \Xi \upsilon_1) \left(2 \left(p_c^+\right)^2 + q_c^- r_c^-\right) + \Xi \upsilon_1 \left(p_c^+\right)^2 \right) &&\notag\\*
	&- 16 \Xi \mathfrak{u}_3 q_c^- \sin\theta &&\notag\\*
	&\quad\times \left(- 4 \sinh\theta_1 \cosh\theta_2 p_c^0 p_c^+ + 3 \sin\theta \left(p_c^+\right)^2 + \sin\theta q_c^- r_c^-\right), &&\\
	\mathfrak{j}_{34,c} =& p_c^+ \left(q_c^-\right)^2 ,&&
\end{flalign}

\begin{flalign}
	\bm{\iota}_4 =& \frac{k^4}{4096 \mathcal{X}^2 \mathcal{Y}^2} \mathfrak{u}_1 \left\{\iota_{40} + \iota_{42} e^{2i\Phi} +  \iota_{44} e^{4i\Phi}\right\}_+ ,&&\notag\\
	\iota_{40} =&  \mathcal{X}^4 \mathfrak{j}_{40,b}  ,&&\notag\\
	\iota_{42} =& 2i^{s+1} \left(\mathfrak{j}_{42,b} \mathcal{X}^4 + (-1)^s \mathfrak{j}_{42,a}^\star \mathcal{Y}^4\right) ,&&\notag\\
	\iota_{44} =& (-1)^{s+1} \left(\mathfrak{j}_{44,b} \mathcal{X}^4 + \mathfrak{j}_{44,a}^\star \mathcal{Y}^4\right), &&\\
	\mathfrak{j}_{40,c} =& \left(p_c^+\right)^4 + 4 \left(p_c^+\right)^2 q_c^- r_c^- + \left(q_c^- r_c^-\right)^2 ,&&\\
	\mathfrak{j}_{42,c} =& i^s p_c^+ q_c^- \left( \left(p_c^+\right)^2 + q_c^- r_c^-\right) ,&&\\
	\mathfrak{j}_{44,c} =& \left(p_c^+ q_c^-\right)^2 ,&&
\end{flalign}
with further abbreviations
\begin{flalign}
	\upsilon_1 =& 16\left( \cos\theta \cosh\theta_1 \cosh\theta_2 + 4 \sin^2\theta - 4 \cosh^2\theta_2\right) ,&& \notag\\
	\upsilon_2 =& 16 \left(1 - \cos\theta \cosh\theta_1 \cosh\theta_2 - \sin^2\theta +\sinh^2\theta_2\right) ,&&\notag\\
	\upsilon_3=& -8 \Bigl(1 -2 \cos\theta \cosh\theta_1 \cosh\theta_2 - 3 \sin^2\theta &&\notag\\
	&\bralign{-8}+ \cosh\left(\theta_1 + \theta_2\right)\cosh\left(\theta_1 - \theta_2\right)\Bigr) ,&&\notag\\
	\upsilon_4 =& -16 \Bigl(1-3 \cos\theta \cosh\theta_1 \cosh\theta_2 - 5 \sin^2\theta - \sinh^2\theta_2 &&\notag\\*
	&\bralign{-16}+4 \cosh\left(\theta_1 + \theta_2\right)\cosh\left(\theta_1 - \theta_2\right) \Bigr) ,&&\notag\\
	\upsilon_5 =& -16 \left(-4 \sin^2\theta + \cosh^2\theta_1 - 3 \cos\theta \cos\theta_1\cos\theta_2+ 2 \cosh^2\theta_2\right)  ,&&\notag\\
	\upsilon_6=& -8 \left(-1 + \cos^2\theta - 2 \cos\theta\cosh\theta_1\cosh\theta_2+ 2 \cosh^2\theta_2\right) ,&&\notag\\
	\upsilon_7 =& -4 \left(-11 \sin^2\theta +\cosh^2\theta_1- 8 \cos\theta \cosh\theta_1 \cosh\theta_2 + 7 \cosh^2\theta_2\right)  ,&&\notag\\
	\upsilon_8 =& 32\Bigl( 2\cos\theta \cosh\theta_1 \cosh\theta_2 - \sinh\left(\theta_1 + \theta_2\right) \sinh\left(\theta_1 - \theta_2\right) + 8 \sin^2\theta &&\notag\\*
	&\bralign{32}- 8 \cosh^2\theta_2\Bigr) ,&&\notag\\
	\chi_1 =& 16 \sin\theta \cosh\theta_1 \sinh\theta_2 ,&&\notag\\
	\chi_2 =& 16 \sin\theta \cosh\theta_2 \sinh\theta_1.&&
\end{flalign}

\section{Coefficients of the Bessel Functions for the Distribution of Cross Sections in GOE systems}
\label{app:coefficientsCross}
Similar, to the scattering matrix elements we have that when exchanging $a\leftrightarrow b$ and setting $\Phi \to -\Phi$, we have $\mathcal{X}(\bm{k}) \to - \mathcal{Y}(\bm{k})$ and vice versa. To save some space, we do not explicitly write the $\bm{k}$ dependence, as it is clear that $\omega, \mathcal{X}, \mathcal{Y}$ are defined by \cref{eqn:XYCrossSections}.
Here, we use a slightly altered definition of the brackets compared to ones for the scattering matrix elements
\begin{equation}
	\{f(a,b,\Phi)\}_+ = f(a,b,\Phi) + f(b,a,-\Phi) .
\end{equation}
The coefficients are
\begin{flalign}
	\bm{\iota}_0 =& \frac{1}{4096} \left(\iota_{00}+\left\{  \iota_{02} e^{2i\Phi} + \iota_{04} e^{4i\Phi} + \iota_{06} e^{6i\Phi} + \iota_{08} e^{8i\Phi}\right\}_+ \right), &&\\
	\iota_{00} =& 2\absk^2
	\Biggl(
	\mathfrak{u}_1 
	\Biggl[
	-16\left(7 p_a^+ p_b^+ +16 p_a^- p_b^- \right) - 64 \Xi^2 p_a^+ p_b^+ \left(\upsilon_1^2 + \chi_1^2\right) &&\notag\\
	&\bralign{2\absk^2}\bralign{\mathfrak{u}_1}+ 64 \Xi \left(8 p_a^- \left(-p_b^0 \chi_1 + 2 p_b^- \upsilon_1\right) - p_a^+ p_b^+ \upsilon_2\right) &&\notag\\
	&\bralign{2\absk^2}\bralign{\mathfrak{u}_1}+ 32 p_a^0 \left(-16\Xi \chi_1 p_b^-  + p_b^0 \left(96 + \absk^2 p_a^+ p_b^+ - 32\Xi \upsilon_1 \right)\right) &&\notag\\
	&\bralign{2\absk^2}\bralign{\mathfrak{u}_1}+ \left(\bmkstar\right)^2 \Biggl(4 p_a^- p_b^- \left(-2 r_a^+ r_b^+ + r_a^- r_b^- \right) &&\notag\\
	&\bralign{2\absk^2}\bralign{\mathfrak{u}_1}\bralign{+\left(\bmkstar\right)^2}+ 2 p_a^+ p_b^+ \left(2r_a^+ r_b^+- r_a^- r_b^-\right)\Biggr)  &&\notag\\
	&\bralign{2\absk^2}\bralign{\mathfrak{u}_1}+ \bmk^2 
	\Biggl(
	8\left(4 \left(p_b^0\right)^2 + \left(p_b^-\right)^2 + q_b^+ r_b^+\right)\left(4 \left(p_a^0\right)^2 + \left(p_a^-\right)^2\right)&&\notag\\
	&\bralign{2\absk^2}\bralign{\mathfrak{u}_1}\bralign{+\bmk^2}+ 8\left(4 \left(p_b^0\right)^2 + \left(p_b^-\right)^2 + q_b^+ r_b^+\right) q_a^+ r_a^+ &&\notag\\
	&\bralign{2\absk^2}\bralign{\mathfrak{u}_1}\bralign{+\bmk^2}+ \left(3 \left(p_b^+\right)^2 + 2 q_b^- r_b^-\right)\left(p_a^+\right)^2 &&\notag\\
	&\bralign{2\absk^2}\bralign{\mathfrak{u}_1}\bralign{+\bmk^2}+2 \left(p_b^+\right)^2 q_a^- r_a^- &&\notag\\
	&\bralign{2\absk^2}\bralign{\mathfrak{u}_1}\bralign{+\bmk^2}+4\left(- 2 p_a^- p_b^- + p_a^+ p_b^+ \right) q_a^+ q_b^+ &&\notag\\
	&\bralign{2\absk^2}\bralign{\mathfrak{u}_1}\bralign{+\bmk^2}+2\left(2 p_a^- p_b^- - p_a^+ p_b^+ -2 r_a^+ r_b^+\right) q_a^- q_b^- &&\notag\\
	&\bralign{2\absk^2}\bralign{\mathfrak{u}_1}\bralign{+\bmk^2}+ 2\left(q_a^- q_b^- - 2 q_a^+ q_b^+\right) r_a^- r_b^- &&\notag\\
	&\bralign{2\absk^2}\bralign{\mathfrak{u}_1}\bralign{+\bmk^2}- 4\left(p_a^+ p_a^- + q_a^+ r_a^- + q_a^- r_a^+\right) p_b^- p_b^+ &&\notag\\
	&\bralign{2\absk^2}\bralign{\mathfrak{u}_1}\bralign{+\bmk^2}- 4 \left(p_b^- p_b^+ + q_b^+ r_b^- + q_b^- r_b^+\right) p_a^+ p_a^- 
	\Biggr)
	\Biggr] &&\notag\\
	&\bralign{2\absk^2}- 64 \mathfrak{u}_3 
	\Biggl[
	\Xi 
	\Biggl(
	p_b^0 \left(-8 p_a^- \chi_1 + 7 p_a^+ \chi_2\right) + p_a^0 \left(-8 p_b^- \chi_1 + 7 p_b^+ \chi_2\right) &&\notag\\
	&\bralign{2\absk^2}\bralign{-64\mathfrak{u}_3}\bralign{\Xi}- p_a^+ p_b^+ \left(16\sin^2\theta \left(-3+2\upsilon_1\right) - \chi_2^2 + \chi_1^2\right) &&\notag\\
	&\bralign{2\absk^2}\bralign{-64\mathfrak{u}_3}\bralign{\Xi}+ 64 \sin^2\theta p_a^- p_b^- 
	\Biggr) &&\notag\\
	&\bralign{2\absk^2}\bralign{-64\mathfrak{u}_3}+ 8 p_a^0 p_b^0 \left(16 - 4\Xi \upsilon_7 + \Xi \left(\upsilon_1^2 + \chi_1^2\right)\right)
	\Biggr] &&\notag\\
	&\bralign{2\absk^2}- 512 \Xi \mathfrak{u}_5 
	\Biggl[
	p_a^0 p_b^0 \left(160 \sin^2\theta - \Xi \left(8 \kappa_5 \upsilon_1 - \chi_2^2 + \chi_1^2\right) \right) &&\notag\\
	&\bralign{2\absk^2}\bralign{-512\Xi\mathfrak{u}_5}+ 32 \Xi \sin^2\theta p_a^+ p_b^+
	\Biggr] &&\notag\\
	&\bralign{2\absk^2}- 131072 \Xi^2 \mathfrak{u}_7 p_a^0 p_b^0 \sin^4\theta
	\Biggr),&&\\
	\iota_{02} =& -2i \absk 
	\Biggl(
	\mathfrak{u}_1 
	\Biggl[
	\absk^2 \bmk 
	\Biggl(
	p_a^+ q_b^-\left(16 p_a^0  p_b^0 + 3 p_a^+ p_b^+\right) &&\notag\\*
	&\bralign{-2i\absk}\bralign{\mathfrak{u}_1}\bralign{\absk^2\bmk}+ 8 q_b^+ p_b^- \left(4 \left(p_a^0\right)^2 + \left(p_a^-\right)^2  \right)&&\notag\\*
	&\bralign{-2i\absk}\bralign{\mathfrak{u}_1}\bralign{\absk^2\bmk}- 4 p_a^+ p_a^- \left(p_b^+ q_b^+ +p_b^- q_b^- \right) &&\notag\\*
	&\bralign{-2i\absk}\bralign{\mathfrak{u}_1}\bralign{\absk^2\bmk}- 2\left(p_b^+ r_a^- - 2 p_b^-r_a^+\right)\left(2 q_a^+ q_b^+ - q_a^- q_b^-\right) 
	\Biggr) &&\notag\\
	&\bralign{-2i\absk}\bralign{\mathfrak{u}_1}- \absk^2\bmkstar 
	\Biggl(
	p_b^+ r_a^-\left( 16 p_a^0 p_b^0 + 3 p_a^+ p_b^+\right) &&\notag\\
	&\bralign{-2i\absk}\bralign{\mathfrak{u}_1}\bralign{-\absk^2\bmkstar}+ 8 p_a^- r_a^+ \left(4 \left(p_b^0\right)^2 + \left(p_b^-\right)^2 \right)&&\notag\\
	&\bralign{-2i\absk}\bralign{\mathfrak{u}_1}\bralign{-\absk^2\bmkstar}- 4 p_b^+ p_b^- \left(p_a^+ r_a^+ + p_a^- r_a^-\right) &&\notag\\
	&\bralign{-2i\absk}\bralign{\mathfrak{u}_1}\bralign{-\absk^2\bmkstar}+ 2\left(p_a^+ q_b^- + 2 p_a^- q_b^+\right)\left(2r_a^+ r_b^+ - r_a^- r_b^-\right)
	\Biggr) &&\notag\\
	&\bralign{-2i\absk}\bralign{\mathfrak{u}_1}+ 16\bmk 
	\Biggl(
	-6 p_a^- q_b^+ +\Xi
	\Biggl(
	4 q_b^+ \left(- 2 p_a^0 \chi_1 + p_a^- \upsilon_1\right) &&\notag\\
	&\bralign{-2i\absk}\bralign{\mathfrak{u}_1}\bralign{+16\bmk}\bralign{-6p_a^- q_b^+ + \Xi}+ p_a^+ q_b^- \left(\upsilon_4 - 2 \Xi \left(\upsilon_1^2 + \chi_1^2\right)\right)
	\Biggr)
	\Biggr) &&\notag\\
	&\bralign{-2i\absk}\bralign{\mathfrak{u}_1}+ 16\bmkstar 
	\Biggl(
	6 r_a^+ p_b^- + \Xi 
	\Biggl(
	4r_a^+ \left(2  p_b^0 \chi_1- p_b^- \upsilon_1\right) &&\notag\\
	&\bralign{-2i\absk}\bralign{\mathfrak{u}_1}\bralign{+16\bmkstar}\bralign{6 r_a^+ p_b^- + \Xi}+ r_a^- p_b^+  \left(-\upsilon_4 + 2 \Xi \left(\upsilon_1^2 + \chi_1^2\right)\right)
	\Biggr)
	\Biggr)
	\Biggr] &&\notag\\
	&+ 32 \Xi \mathfrak{u}_3 
	\Biggl[
	2\bmk\left( p_a^0 \left(-q_b^- \chi_2  + 2 q_b^+ \chi_1 \right) - 4\sin^2\theta\left(4  p_a^- q_b^+ + 7 p_a^+ q_b^-\right) \right) &&\notag\\
	&\bralign{+32\Xi\mathfrak{u}_3}+ 2\bmkstar\left(p_b^0\left( r_a^- \chi_2  - 2 r_a^+ \chi_1\right) + 4\sin^2\theta\left(4 \bmkstar p_b^- r_a^+ + 7 p_b^+ r_a^-\right)\right) &&\notag\\
	&\bralign{+32\Xi\mathfrak{u}_3}+ \Xi \left(32 \sin^2\theta \upsilon_1 - \chi_2^2 + \chi_1^2\right) \left(\bmk p_a^+ q_b^- - \bmkstar p_b^+ r_a^-\right)
	\Biggr] &&\notag\\
	&+ 8192 \Xi^2 \mathfrak{u}_5 \sin^4\theta \left(-\bmk p_a^+ q_b^- + \bmkstar p_b^+ r_a^-\right)
	\Biggr),&&\\
	\iota_{04} =& \absk^2 
	\Biggl(
	-\mathfrak{u}_1 
	\Biggl[
	\left(\bmk p_a^+ q_b^-\right)^2  + \left(\bmkstar p_b^+ r_a^-\right)^2 &&\notag\\*
	&\bralign{\absk^2}\bralign{-\mathfrak{u}_1}+ 8\absk^2
	\Biggl( 
	p_a^-r_a^- p_b^+ q_b^+ - 2 p_a^- r_a^+ p_b^- q_b^+  &&\notag\\*
	&\bralign{\absk^2}\bralign{-\mathfrak{u}_1}\bralign{+8\absk^2}+ p_a^+ r_a^+ p_b^- q_b^- + p_a^+ r_a^- p_b^+ q_b^-
	\Biggr) &&\notag\\*
	&\bralign{\absk^2}\bralign{-\mathfrak{u}_1} + 32 q_b^- r_a^- \left(2 - \absk^2 p_a^0 p_b^0 + 2 \Xi \left(-2 \upsilon_3 + \Xi \left(\upsilon_1^2 + \chi_1^2\right)\right)\right)
	\Biggr] &&\notag\\*
	&\bralign{\absk^2}+ 64  \Xi \mathfrak{u}_3 q_b^- r_a^- \left(-64 \sin^2\theta + \Xi \left(32 \upsilon_1 \sin^2\theta- \chi_2^2 + \chi_1^2\right)\right) &&\notag\\*
	&\bralign{\absk^2}- 16384 \Xi^2 \mathfrak{u}_5 q_b^- r_a^- \sin^4\theta
	\Biggr),&&\\
	\iota_{06} =& -2i \absk^3 \mathfrak{u}_1 q_b^- r_a^-  \left(\bmk p_a^+ q_b^- - \bmkstar p_b^+ r_a^-\right),&&\\
	\iota_{08} =& \mathfrak{u}_1 \left(\absk^2 q_b^- r_a^-\right)^2 ,&&
\end{flalign}

\begin{flalign}
	\bm{\iota}_1 =& -\frac{i}{1024 \sqrt{\mathcal{X} \mathcal{Y}}} \left\{\iota_{10} + \iota_{12} e^{2i\Phi} + \iota_{14} e^{4i\Phi} + \iota_{16} e^{6i\Phi}\right\}_+ ,&&\notag\\
	\iota_{10} =& 2 i \absk \mathcal{X} \mathfrak{j}_{10,ba},&&\notag\\
	\iota_{12} =& 2 \absk^2 \left(\mathfrak{j}_{12,ba} \mathcal{X} + \mathfrak{j}_{12,ab}^\star \mathcal{Y}\right),&&\notag\\
	\iota_{14} =& i \left(\mathfrak{j}_{14,ba} \mathcal{X} - \mathfrak{j}_{14,ab}^\star \mathcal{Y}\right),&&\notag\\
	\iota_{16} =& 2 \left(\mathfrak{j}_{16,ba} \mathcal{X} + \mathfrak{j}_{16,ab}^\star \mathcal{Y}\right),&&\\
	\mathfrak{j}_{10,c c\myprime} =&\mathfrak{u}_1 
	\Biggl[
	\absk^2 
	\Biggl(
	8 p_a^0 \left(\Xi\chi_1 \left(2 p_b^+ p_b^- + q_b^+ r_b^- + q_b^+ r_b^-\right) + 2 p_b^0 p_b^+ \left(-3 + 2 \Xi \upsilon_1\right)\right) &&\notag\\*
	&\bralign{\mathfrak{u}_1}\bralign{\absk^2}+ p_a^+ 
	\Biggl(
	32 \left(p_b^0\right)^2 - 3 \left(p_b^+\right)^2 + 8 \left(p_b^-\right)^2 + 8 q_b^+ r_b^+ - 2 q_b^- r_b^- &&\notag\\*
	&\bralign{\mathfrak{u}_1}\bralign{\absk^2}\bralign{+p_a^+}+ 2 \Xi \upsilon_1 \left( 3 \left(p_b^+\right)^2 + 2 q_b^- r_b^-\right)
	\Biggr) &&\notag\\*
	&\bralign{\mathfrak{u}_1}\bralign{\absk^2}- 4 p_a^- \Biggl( \Xi\left( -4 \chi_1  p_b^0 p_b^+ + 2 \upsilon_1 p_b^- p_b^+ +  \upsilon_1 \left(q_b^+ r_b^- + q_b^- r_b^+\right)\right) &&\notag\\*
	&\bralign{\mathfrak{u}_1}\bralign{\absk^2}\bralign{-4p_a^-} +p_b^- p_b^+\Biggr)
	\Biggr) &&\notag\\*
	&+ p_b^+ 
	\Biggl(
	\bmk^2 \left(q_a^- q_b^- - 2 q_a^+ q_b^+\right) \left(-3 + 2 \Xi \upsilon_1\right) &&\notag\\*
	&\bralign{\mathfrak{u}_1}\bralign{\mathfrak{p_c^+}}+ \left(\bmkstar\right)^2 \left(r_a^- r_b^- - 2 r_a^+ r_b^+\right) \left(-3 + 2 \Xi \upsilon_1\right) &&\notag\\*
	&\bralign{\mathfrak{u}_1}\bralign{\mathfrak{p_c^+}}+ 8 \left(9 + 4 \Xi \left(- \upsilon_5 + \Xi \left(\upsilon_1^2 + \chi_1^2\right)\right)\right)
	\Biggr)
	\Biggr] &&\notag\\*
	&+ 8 \Xi \mathfrak{u}_3 
	\Biggl[
	-640 p_b^+ p_b^0\chi_2 \sin^2\theta  \left(-8 + \absk^2 \left(8 p_a^0 p_b^0 +p_a^+ p_b^+\right)\right) &&\notag\\*
	&\bralign{+4\Xi \mathfrak{u}_3} -320 \absk^2 p_a^0 p_b^+ \chi_2 \sin^2\theta &&\notag\\*
	&\bralign{+4\Xi \mathfrak{u}_3}\quad\times\left(4 \left(p_b^-\right)^2 + \left(p_b^+\right)^2 + q_b^- r_b^- + 4 q_b^+ r_b^+\right)\Biggr) &&\notag\\*
	&\bralign{+4\Xi \mathfrak{u}_3}- \chi_1 \absk^2 \left(2 p_a^- p_b^0 p_b^+ + p_a^0 \left(2 p_b^- p_b^+ + q_b^+ r_b^- + q_b^- r_b^+\right)\right)&&\notag\\*
	&\bralign{+4\Xi \mathfrak{u}_3}- 4 \absk^2 \sin^2\theta 
	\Biggl(
	16 p_a^0 p_b^0 p_b^+ +  p_a^+ \left(3 \left(p_b^+\right)^2 + 2 q_b^- r_b^-\right) &&\notag\\*
	&\bralign{+4\Xi \mathfrak{u}_3}\bralign{-4\absk^2\sin^2\theta}- 2 p_a^- \left(2p_b^- p_b^+ + q_b^+ r_b^- + q_b^- r_b^+\right)
	\Biggr) &&\notag\\*
	&\bralign{+4\Xi \mathfrak{u}_3}+4 p_b^+ \sin^2\theta
	\Biggl( 
	\bmk^2 \left(q_a^- q_b^- - 2 q_a^+ q_b^+\right) &&\notag\\*
	&\bralign{+4\Xi \mathfrak{u}_3}\bralign{+4p_b^+\sin^2\theta}+ \left(\bmkstar\right)^2 \left(r_a^- r_b^- - 2 r_a^+ r_b^+\right)
	\Biggr) &&\notag\\*
	&\bralign{+4\Xi \mathfrak{u}_3}+4 p_b^+ \Xi \left(8 \kappa_5 \upsilon_1 + \chi_1^2 - \chi_2^2\right)
	\Biggr] &&\notag\\*
	&- 8192 \Xi^2 \mathfrak{u}_5 p_b^+ \sin^4\theta ,&&\\
	\mathfrak{j}_{12,c c\myprime} =& \mathfrak{u}_1 \Biggl[
	2 \bmk \Biggl(
	2q_c^- \left(- p_c^+ p_{c\myprime}^+ - p_c^- p_{c\myprime}^- + 2 \Xi \chi_1 p_c^0 q_c^- p_{c\myprime}^- \right) &&\notag\\*
	&\bralign{\mathfrak{u}_1}\bralign{2\bmk}+ q_c^+ \left(4 p_c^- p_{c\myprime}^+  + p_c^+ p_{c\myprime}^-\right)  &&\notag\\*
	&\bralign{\mathfrak{u}_1}\bralign{2\bmk}+ \Xi \upsilon_1 \left(3 p_{c\myprime}^+ p_c^+ q_c^- - 2p_{c\myprime}^- \left(p_c^- q_c^- + p_c^+ q_c^+\right)\right) &&\notag\\*
	&\bralign{\mathfrak{u}_1}\bralign{2\bmk}+ 4 p_{c\myprime}^0 \left(\Xi\chi_1 \left(p_c^- q_c^- + p_c^+ q_c^+\right) + 2 p_c^0 q_c^- \left(-2 + \Xi \upsilon_1\right)\right)
	\Biggr)&&\notag\\*
	&+ \bmkstar \Biggl(
	2 q_c^- \left(2 r_c^+ r_{c\myprime}^+ - r_c^- r_{c\myprime}^-\right) \left(-1 + \Xi \upsilon_1\right) - \left(p_c^+\right)^2 r_{c\myprime}^- \left(2 - 3 \Xi \upsilon_1\right) &&\notag\\*
	&\bralign{\mathfrak{u}_1}\bralign{+\bmkstar}+ 2 p_c^+ r_{c\myprime}^+ \left(-4\Xi \chi_1 p_c^0  + p_c^- \left(-1 + 2 \Xi \upsilon_1\right)\right)
	\Biggr)
	\Biggr] &&\notag\\*
	&-4\Xi \mathfrak{u}_3 
	\Biggl[
	\bmk 
	\Biggl(
	8\sin^2\theta \left(q_c^-\left(8 p_c^0 p_{c\myprime}^0 + 3 p_c^+ p_{c\myprime}^+ \right) - 2 p_{c\myprime}^- \left(p_c^- q_c^- + p_c^+ q_c^+\right)\right) &&\notag\\*
	&\bralign{-4\Xi \mathfrak{u}_3}\bralign{\bmk}+ 2 \chi_1 \left(q_c^-\left( p_c^0 p_{c\myprime}^- + p_c^-  p_{c\myprime}^0\right) + p_c^+ q_c^+ p_{c\myprime}^0\right) &&\notag\\*
	&\bralign{-4\Xi \mathfrak{u}_3}\bralign{\bmk}- \chi_2  \left(q_c^-\left( p_c^0 p_{c\myprime}^+ + p_c^+ p_{c\myprime}^0 \right) + 4 p_c^- q_c^+ p_{c\myprime}^0\right) 
	\Biggr) &&\notag\\*
	&\bralign{-4\Xi\mathfrak{u}_3}+ \bmkstar 
	\Biggl(
	4\sin^2\theta\Biggl(4 r_{c\myprime}^+ \left(p_c^- p_c^+ + q_c^- r_c^+\right) &&\notag\\*
	&\bralign{-4\Xi\mathfrak{u}_3}\bralign{+\bmkstar}\bralign{4\sin^2\theta}-r_{c\myprime}^- \left(3 \left(p_c^+\right)^2 + 2 q_c^- r_c^- \right) \Biggr) &&\notag\\*
	&\bralign{-4\Xi\mathfrak{u}_3}\bralign{+\bmkstar}+p_c^0 \left(\chi_2  p_c^+ r_{c\myprime}^- - 2 \chi_1 p_c^+ r_{c\myprime}^+\right)
	\Biggr)
	\Biggr],&&\\
	\mathfrak{j}_{14,c c\myprime} =&  - \absk \mathfrak{u}_1q_c^- 
	\Biggl[
	4 \absk^2 \left(p_c^+ r_{c\myprime}^- - 2\Xi \left(2 \chi_1 p_c^0 r_{c\myprime}^+ + \upsilon_1 \left(p_c^+ r_{c\myprime}^- - p_c^- r_{c\myprime}^+\right)\right)\right) &&\notag\\
	&\bralign{- \absk \mathfrak{u}_1q_c^-} +\bmk^2 q_c^- p_{c\myprime}^+ \left(-3 + 2\Xi \upsilon_1\right)
	\Biggr] &&\notag\\
	&+ 8 \Xi \absk  \mathfrak{u}_3 q_c^-
	\Biggl[
	\absk^2 p_c^0 \left(\chi_2 r_{c\myprime}^- - 2 \chi_1 r_{c\myprime}^+\right) + 4 \bmk^2 \sin^2\theta p_{c\myprime}^+ q_c^- &&\notag\\
	&\bralign{+8\Xi \absk \mathfrak{u}_3 q_c^-}+ 16 \absk^2 \sin^2\theta \left(-p_c^+ r_{c\myprime}^- + p_c^- r_{c\myprime}^+\right)
	\Biggr],&&\\
	\mathfrak{j}_{16,c c\myprime} =& \left(\absk q_c^-\right)^2 r_{c\myprime}^- \left(\mathfrak{u}_1 \left(-1 + \Xi \upsilon_1\right) - 4 \Xi \kappa_5 \mathfrak{u}_3\right),&&
\end{flalign}

\begin{flalign}
	\bm{\iota}_2 =& \frac{1}{2048 \mathcal{X} \mathcal{Y}} \left\{\iota_{20} + \iota_{22} e^{2i\Phi} + \iota_{24} e^{4i\Phi} + \iota_{26} e^{6i\Phi}\right\}_+ ,&&\notag\\
	\iota_{20} =& \mathcal{X}^2 \absk^2 \mathfrak{j}_{20,ba},&&\notag\\
	\iota_{22} =& i\absk \left(\mathfrak{j}_{22,ba} \mathcal{X}^2 - \mathfrak{j}_{22,ab}^\star \mathcal{Y}^2\right),&&\notag\\
	\iota_{24} =& \left(\mathfrak{j}_{24,ba} \mathcal{X}^2 - \mathfrak{j}_{24,ab}^\star \mathcal{Y}^2\right),&&\notag\\
	\iota_{26} =& i \left(\mathfrak{j}_{26,ba} \mathcal{X}^2 - \mathfrak{j}_{26,ab}^\star \mathcal{Y}^2\right),&&\\
	\mathfrak{j}_{20,c c\myprime} =& \mathfrak{u}_1 
	\Biggl[
	4 \absk^2 p_c^+ \left(p_{c\myprime}^+ q_c^- r_c^- + p_{c\myprime}^- \left(q_c^+ r_c^- + q_c^- r_c^+\right)\right) &&\notag\\*
	&\bralign{\mathfrak{u}_1}+ 16 q_c^- r_c^- \left(2 - \absk^2 p_{c\myprime}^0 p_c^0 +2 \Xi\left(-  \upsilon_3 + \Xi \left(\upsilon_1^2 + \chi_1^2\right)\right)\right)&&\notag\\*
	&\bralign{\mathfrak{u}_1}+ \left(p_c^+\right)^2 
	\Biggl(
	-8 + 2 \absk^2 \left(-8 p_{c\myprime}^0 p_c^0 - p_{c\myprime}^+ p_c^+ + 2p_{c\myprime}^- p_c^-\right)  &&\notag\\*
	&\bralign{\mathfrak{u}_1}\bralign{+\left(p_c^+\right)^2}+ \bmk^2 \left(q_{c\myprime}^- q_c^- - 2q_{c\myprime}^+ q_c^+\right) + \left(\bmkstar\right)^2 \left(r_{c\myprime}^- r_c^- - 2 r_{c\myprime}^+ r_c^+\right) &&\notag\\*
	&\bralign{\mathfrak{u}_1}\bralign{+\left(p_c^+\right)^2}+ 32 \Xi \left(\upsilon_2 + \Xi \left(\upsilon_1^2 + \chi_1^2\right)\right)
	\Biggr)
	\Biggr] &&\notag\\*
	&+ 32 \Xi \mathfrak{u}_3 \Biggl[ 6 \chi_2 p_c^0 p_b^+ + 16 \sin^2\theta\left(3 \left(p_c^+\right)^2 + + q_c^- r_c^-\right) &&\notag\\*
	&\bralign{+32 \Xi \mathfrak{u}_3}- \Xi \left(\left(p_c^+\right)^2 + q_c^- r_c^-\right) \left(32 \sin^2\theta \upsilon_1 - \kappa_7^2 + \kappa_8^2\right)\Biggr] &&\notag\\*
	&+ 8192 \Xi^2 \mathfrak{u}_5 \left(\left(p_c^+\right)^2 + q_c^- r_c^-\right) \sin^4\theta ,&&\\
	\mathfrak{j}_{22,c c\myprime} =& \mathfrak{u}_1 \Biggl[
	\absk^2 \Biggl(
	\bmkstar p_c^+ \left(p_c^+\left(- p_c^+ r_{c\myprime}^- + 2 p_c^- r_{c\myprime}^+\right) + 2 q_c^-\left(2 q_c^- r_{c\myprime}^+ r_c^+ -  r_{c\myprime}^- r_c^-  \right)\right) &&\notag\\
	&\bralign{\mathfrak{u}_1}\bralign{\absk^2}+ \bmk \Biggl(4  p_c^+ q_c^- \left(4 p_c^0 p_{c\myprime}^0 - p_c^- p_{c\myprime}^- \right) &&\notag\\
	&\bralign{\mathfrak{u}_1}\bralign{\absk^2}\bralign{+\bmk}+ p_{c\myprime}^+ q_c^- \left(3 \left(p_c^+\right)^2 + q_c^- r_c^-\right) &&\notag\\
	&\bralign{\mathfrak{u}_1}\bralign{\absk^2}\bralign{+\bmk}- 2 p_{c\myprime}^- \left(+ \left(p_c^+\right)^2 q_c^+ + \left(q_c^-\right)^2 r_c^+\right)\Biggr)
	\Biggr) &&\notag\\
	&+ 16 \bmk p_c^+ q_c^- \Xi \left(\upsilon_4 - 2 \Xi \left(\upsilon_1^2 + \chi_1^2\right)\right)
	\Biggr]&&\notag\\
	&- 32 \Xi\bmk \mathfrak{u}_3 q_c^- \left(2 \left(\chi_2 p_c^0 + 28 \sin^2\theta p_c^+\right) - \Xi p_c^+ \left(32 \sin^2\theta \upsilon_1 - \chi_2^2 + \chi_1^2\right)\right) &&\notag\\
	&- 8192 \Xi^2 \bmk \mathfrak{u}_5 q_c^- p_c^+ \sin^4\theta ,&&\\
	\mathfrak{j}_{24,c c\myprime} =& \absk^2 q_c^- \mathfrak{u}_1 \Biggl(
	\bmk^2 q_c^- \left( p_{c\myprime}^+ p_c^+-2 p_{c\myprime}^- p_c^- \right) &&\notag\\
	&\bralign{\absk^2 q_c^- \mathfrak{u}_1}+ \absk^2 \left(-r_{c\myprime}^-\left(2 \left(p_c^+\right)^2 + q_c^-  r_c^-\right) + 2r_{c\myprime}^+ \left(2 p_c^- p_c^+ + q_c^- r_c^+\right)\right)
	\Biggr),&&\\
	\mathfrak{j}_{26,c c\myprime} =& \bmk \absk^3  \left(q_c^-\right)^2 \mathfrak{u}_1 \left(p_c^+ r_{c\myprime}^- - 2 p_c^- r_{c\myprime}^+\right),&&
\end{flalign}

\begin{flalign}
	\bm{\iota}_3 =& i\frac{\sqrt{\mathcal{X} \mathcal{Y}}}{1024 \mathcal{X}^2 \mathcal{Y}^2} \left\{\iota_{30} + \iota_{32} e^{2i\Phi} + \iota_{34} e^{4i\Phi}\right\}_+ ,&&\notag\\
	\iota_{30} =& 2 i \mathcal{X}^3 \absk^3 \mathfrak{j}_{30,b} ,&&\notag\\
	\iota_{32} =& 2 \left(\mathfrak{j}_{32,b} \mathcal{X}^3 + \mathfrak{j}_{32,a}^\star \mathcal{Y}^3\right), &&\notag\\
	\iota_{34} =& i \left(\mathfrak{j}_{34,b} \mathcal{X}^3 - \mathfrak{j}_{34,a}^\star \mathcal{Y}^3\right) ,&&\\
	\mathfrak{j}_{30,c} =& p_c^+ \left(-1 + 2 \Xi \upsilon_1\right) \mathfrak{u}_1 \left(2 q_c^- r_c^- + \left(p_c^+\right)^2\right)&&\notag\\*
	&+ 4 \Xi \mathfrak{u}_3 \left(\chi_2 p_c^0  \left(q_c^- r_c^- \left(p_c^+\right)^2\right) - 8\sin^2\theta p_c^+ \left(2q_c^- r_c^- + \left(p_c^+\right)^2\right)\right) ,&&\\
	\mathfrak{j}_{32,c} =& \bmk \absk^2 q_c^- \Biggl(
	\mathfrak{u}_1 \left(q_c^- r_c^- \left(-1 + \Xi \upsilon_1\right) + \left(p_c^+\right)^2 \left(-2 + 3 \Xi \upsilon_1\right)\right) &&\notag\\*
	&\bralign{\bmk \absk^2 q_c^-}- 4\Xi \mathfrak{u}_3 \left( 4\sin^2\theta q_c^- r_c^- + p_c^+ \left(-\chi_2 p_c^0 + 12 \sin^2\theta p_c^+\right)\right)
	\Biggr) ,&&\\
	\mathfrak{j}_{34,c} =& \absk p_c^+ \left(\bmk q_c^-\right)^2 \left(\mathfrak{u}_1 \left(-3+2\Xi \upsilon_1\right)-32\Xi\mathfrak{u}_3 \sin^2\theta \right)
\end{flalign}

\begin{flalign}
	\bm{\iota}_4 =& \frac{\mathfrak{u}_1 }{4096 \mathcal{X}^2 \mathcal{Y}^2} \left\{\iota_{40} + \iota_{42} e^{2i\Phi} + \iota_{44} e^{4i\Phi}\right\}_+ ,&&\notag\\
	\iota_{40} =& \mathcal{X}^4 \absk^4\left( \left(p_b^+\right)^4 + \left(q_b^- r_b^-\right)^2 + 4 \left(p_b^+\right)^2 q_b^- r_b^-\right) ,&&\notag\\
	\iota_{42} =& 2 i\left(\mathfrak{j}_{42,b} \mathcal{X}^4 - \mathfrak{j}_{42,a}^\star \mathcal{Y}^4\right), &&\notag\\
	\iota_{44} =& \left(\mathfrak{j}_{44,b} \mathcal{X}^4 + \mathfrak{j}_{44,a}^\star \mathcal{Y}^4\right)  ,&&\\
	\mathfrak{j}_{42,c} =& -\bmk \absk^3  p_c^+ q_c^- \left( q_c^- r_c^- + \left(p_c^+\right)^2 \right) ,&&\\
	\mathfrak{j}_{44,c} =& - \left(\bmk \absk p_c^+ q_c^-\right)^2 .&&
\end{flalign}

\section*{References}

\bibliography{bibliography.bib,streu_refs_combined.bib,Lett_Bib.bib}

\bibliographystyle{iopart-num}

\end{document}